\documentclass[usenatbib,useAMS]{mn2e}

\usepackage{epsfig}
\usepackage{amsmath}
\usepackage{multirow}
\usepackage[english]{babel}
\usepackage{url}
\usepackage{amssymb}
\usepackage{graphicx}
\usepackage{graphics}
\usepackage{deluxetable}
\usepackage{lscape}
\usepackage{rotating}
\usepackage{longtable}
\usepackage{natbib}

\title[The hosts of SLSNe]{Spectroscopy of superluminous supernova host galaxies. A preference of hydrogen-poor events for extreme emission line galaxies.}

\author[G. Leloudas et al.]{G.~Leloudas$^{1,2}$\thanks{E-mail: giorgos@dark-cosmology.dk}, 
S.~Schulze$^{3,4}$,  
T.~Kr\"{u}hler$^{5}$, 
J.~Gorosabel$^{6,7,8}$,
L.~Christensen$^{1}$,
\newauthor
A.~Mehner$^{5}$, 
A.~de Ugarte Postigo$^{6,1}$,
R.~Amor\'{\i}n$^{9}$,
C.~C. Th\"{o}ne$^{6}$,
J.~P.~Anderson$^{5}$,
\newauthor
F.~E. Bauer$^{3,4,10}$,
A.~Gallazzi$^{11,1}$,
K.~G. He{\l}miniak$^{12,13}$,
J.~Hjorth$^{1}$,
E.~Ibar$^{14}$,
\newauthor
D.~Malesani$^{1}$,
N.~Morrell$^{15}$,
J.~Vinko$^{16,17}$
and J.~C. Wheeler$^{17}$
 \\
$^{1}$Dark Cosmology Centre, Niels Bohr Institute, University of Copenhagen, Juliane Maries Vej 30, 2100 Copenhagen, Denmark.\\
$^{2}$Department of Particle Physics \& Astrophysics, Weizmann Institute of Science, Rehovot 76100, Israel.\\ 
$^{3}$Instituto de Astrof\'{\i}sica, Facultad de F\'{\i}sica, Pontificia Universidad Cat\'{o}lica de Chile, 306, Santiago 22, Chile.\\ 
$^{4}$Millennium Institute of Astrophysics, Vicu\~{n}a Mackenna 4860, 7820436 Macul, Santiago, Chile.\\
$^{5}$ESO, Alonso de Cordova 3107, Vitacura, Santiago de Chile, Chile.\\
$^{6}$Instituto de Astrof\' isica de Andaluc\' ia (IAA-CSIC), Glorieta de la Astronom\' ia s/n, E-18008, Granada, Spain.\\
$^{7}$Unidad Asociada Grupo Ciencia Planetarias UPV/EHU-IAA/CSIC, Departamento de F\'isica Aplicada I, E.T.S. Ingenier\'ia, \\
~Universidad del Pa\'is-Vasco UPV/EHU, Alameda de Urquijo s/n, E-48013 Bilbao, Spain.\\
$^{8}$Ikerbasque, Basque Foundation for Science, Alameda de Urquijo 36-5, E-48008 Bilbao, Spain.\\
$^{9}$INAF $-$ Osservatorio Astronomico di Roma, via Frascati 33, 00040 Monteporzio Catone, Roma, Italy.\\
$^{10}$Space Science Institute, 4750 Walnut Street, Suite 205, Boulder, Colorado 80301, USA.\\
$^{11}$INAF $-$ INAF $-$ Osservatorio Astrofisico di Arcetri, Largo Enrico Fermi 5, 50125 Firenze, Italy.\\
$^{12}$Subaru Telescope, National Astronomical Observatory of Japan, 650 North Aohoku Place, Hilo, HI 96720, USA.\\
$^{13}$Nicolaus Copernicus Astronomical Center, Department of Astrophysics, ul. Rabia\'{n}ska 8, 87-100 Toru\'{n}, Poland.\\
$^{14}$Instituto de F\'isica y Astronom\'ia, Universidad de Valpara\'iso, Avda. Gran Breta\~na 1111, Valpara\'iso, Chile.\\
$^{15}$Las Campanas Observatory, Carnegie Observatories, Casilla 601, La Serena, Chile.\\
$^{16}$Department of Optics and Quantum Electronics, University of Szeged, Szeged, Dem ter 9, 6720, Hungary.\\
$^{17}$Department of Astronomy, University of Texas at Austin, Austin, TX 78712, USA.}

\begin{document}

\date{}

\pagerange{\pageref{firstpage}--\pageref{lastpage}} \pubyear{2002}

\maketitle

\label{firstpage}

\begin{abstract}
Superluminous supernovae (SLSNe) are very bright explosions that were only discovered recently
and that show a preference for occurring in faint dwarf galaxies. 
Understanding why stellar evolution yields different types of stellar explosions in these environments is fundamental in order to both uncover the elusive progenitors of SLSNe and to study star formation in dwarf galaxies. In this paper, we present the first results of our project to study SUperluminous Supernova Host galaxIES, focusing on the sample for which we have obtained spectroscopy. We show that SLSNe-I and SLSNe-R (hydrogen-poor) often ($\sim$$50 \%$ in our sample) occur in a class of galaxies that is known as Extreme Emission Line Galaxies (EELGs). The probability of this happening by chance is negligible and we therefore conclude that the extreme environmental conditions and the SLSN phenomenon are related. In contrast, SLSNe-II (hydrogen-rich) occur in more massive, more metal-rich galaxies with softer radiation fields. Therefore, if SLSNe-II constitute a uniform class, their progenitor systems are likely different from those of H-poor SLSNe. Gamma-ray bursts (GRBs) are, on average, not found in as extreme environments as H-poor SLSNe. We propose that H-poor SLSNe result from the very first stars exploding in a starburst, even earlier than GRBs. This might indicate a bottom-light initial mass function in these systems. SLSNe present a novel method of selecting candidate EELGs independent of their luminosity.
\end{abstract}


\clearpage

\section{Introduction}

Since Zwicky started the first systematic searches for supernovae (SNe), most events were discovered by monitoring nearby bright galaxies.
With few exceptions, the SNe discovered in this traditional way fell within the framework of the classical SN classification scheme \citep[e.g.][]{1997ARA&A..35..309F}.
During the last decade, modern transient surveys \citep[e.g.][]{2005AAS...20717102Q,2009ApJ...696..870D,2009PASP..121.1395L,2012ApJ...750...99T}, which do not target specific galaxies, contributed to the discovery of new classes of transients.
Among these newly established classes, were Superluminous SNe \citep[SLSNe, e.g.][]{2007ApJ...666.1116S,2007ApJ...659L..13O,2007ApJ...668L..99Q}, 
i.e. explosions 10--100 times brighter than ordinary SNe  \cite[for a review, see][]{2012Sci...337..927G}.
To date, we have no good handle of what produces SLSNe, especially those that do not show hydrogen in their spectra.
However, it is reasonable to assume that the environment plays an important role in their formation.
In the opposite case, and despite their low volumetric rate \citep{2013MNRAS.431..912Q}, it is possible that SLSNe might have been discovered earlier in targeted surveys.

The study of SLSN host galaxies is in its infancy. \cite{2011ApJ...727...15N} presented a first study based on a sample of 17 hosts. However, this study was solely based on archival data of two photometric bands (SDSS $r$ and \textit{GALEX} NUV) with a significant number of non-detections, and a sample that contained a few ambiguous objects.
Despite these drawbacks, they concluded that SLSNe preferentially occur in low-mass galaxies with high specific star formation rates (sSFR).
Two individual hosts were studied in detail by \cite{2013ApJ...763L..28C} and \cite{2013ApJ...771...97L}, who showed that their metallicities were low (below 0.1 $Z_{\sun}$).
A systematic study was presented by \cite{2014ApJ...787..138L}  focusing on a sample of 30 SLSN hosts, including 14 objects detected by the PanSTARRS survey.
The main conclusion of their study, including spectroscopy of 12 objects, was that the hosts of H-poor SLSNe (no H-rich events were studied) are similar to Gamma-ray burst (GRB) host galaxies. 
The authors favour a magnetar origin for (H-poor) SLSN explosions. 
Recently, \cite{2014arXiv1409.7728C} and \cite{2014arXiv1411.1104T} studied in detail the host of  PTF12dam and \cite{2014ApJ...797...24V} examined the absorption properties towards H-poor SLSN sight lines. Finally, \cite{2014arXiv1411.1060L} presented a resolved \textit{HST} study of 16 H-poor SLSN hosts.

Here, we present our own programme to study SLSN hosts, a project which we have dubbed SUSHIES.\footnote{from SUperluminous Supernova Host galaxIES.}
In this paper we focus on the sub-sample of SLSN hosts for which we have secured spectroscopy (until 2014 June). 
We study their spectroscopic properties and find important differences between both H-poor and H-rich SLSN hosts but even between H-poor SLSN and GRB hosts.
We show that H-poor SLSNe show a preference  for a class of low-mass, metal poor, intensively star-forming galaxies with strong nebular emission lines and a hard ionizing radiation field, often referred to as Extreme Emission Line Galaxies  \citep[EELGs; e.g.][]{2011ApJ...743..121A,2014arXiv1403.3441A,2014A&A...568L...8A}.
The study of the full SUSHIES sample, as well as a more detailed analysis of the properties derived from imaging observations, will be presented in forthcoming publications (Schulze et al., in prep.). 

Section~\ref{sec:obs} describes our host sample and how the data were collected, reduced and analysed.
Section~\ref{sec:results} presents our results, and in Section~\ref{sec:disc} we discuss the implications of these findings. 
Section~\ref{sec:conc} summarizes our conclusions.
Throughout this paper we adopt a \textit{Planck} Cosmology, i.e. $\Omega_{\rm m} = 0.315, \Omega_{\Lambda} = 0.685, \rm{H}_{0}=67.3$ \citep{2014A&A...571A..16P}. All errors quoted are at the 1$\sigma$ level.

\section{Observations and Methods}
\label{sec:obs}

\subsection{Sample}

In SUSHIES we target all hosts of SLSNe that are publicly announced in telegrams, circulars or refereed publications. 
Before PanSTARRS released their SLSN sample \citep{2014ApJ...787..138L}, the public SLSNe numbered about $\sim$30 events in the redshift range $0.1<z<1.6$, and possibly extending up to $z\sim4$ \citep{2012Natur.491..228C}.
We obtain multi-band photometry for these galaxies using a variety of telescopes, and we complement our data with archival (e.g. SDSS) and literature data, when available.
For spectroscopy, we have been using a variety of instruments on 6--10m telescopes, including FORS2 and X-shooter on the VLT, OSIRIS on the GTC and IMACS on Magellan.
Spectroscopy has been obtained for the majority of hosts brighter then $R = 24\,\rm{mag}$. 
SUSHIES continues to obtain data. This paper contains data obtained up to 2014 June. 
Potential biases arising from the sample selection are discussed in Section~\ref{subsec:bias}.

\begin{table*}
 \centering
 \begin{minipage}{170mm}
  \caption{The SLSN host sample, including a log of the spectroscopic observations. The horizontal line separates data obtained with our programs and data retrieved from archives. Above and below the line, the SNe are listed by discovery year.}
  \label{tab:log}
  \begin{tabular}{@{}lccccccccc@{}}
\hline
SLSN  & Type & Ref. & $z$ & RA & Dec.  & Instrument$^a$ & Resolution$^b$ & Exp. Time & Date \\
            &            &         &    &  (J2000)     &   (J2000)        &                               &  (\AA)            &       (s) 	      &  (UT) \\
\hline
SN 1999as & R  &  [1] & 0.127 & 09:16:30.86 & $+$13:39:02.2 & OSIRIS &  7.1  &  1800 & 2014-02-21\\    
SN 1999bd & II  &  [1]  & 0.151 & 09:30:29.17 & $+$16:26:07.8 & IMACS & 3.4 & 3600 & 2013-02-08\\  
SN 2006oz & I  &  [2]  & 0.396 &  22:08:53.56 & $+$00:53:50.4 & X-shooter &  0.8 & 2700 & 2012-09-15 \\ 
SN 2006tf  & II  &  [3]  & 0.074 & 12:46:15.82 & $+$11:25:56.3 & FORS2 & 11.3 & 2700 & 2013-05-29 \\  
SN 2009jh$^1$  & I  &  [4] & 0.349 & 14:49:10.08 & $+$29:25:11.4 & X-shooter & 1.1 & 5400 & 2012-06-17 \\   
PTF09cnd  &I  &  [4] & 0.258 & 16:12:08.94 & $+$51:29:16.1 & OSIRIS & 13.1 & 4400$^c$ & 2013-05-03 \\  
SN 2010kd & R  &  [5,6] & 0.101 & 12:08:01.11 & $+$49:13:31.1 & OSIRIS & 12.5 & 2700 & 2013-05-17 \\  
PTF10heh    & II  &  [7] & 0.338 & 12:48:52.04 & $+$13:26:24.5&  X-shooter & 0.7 & 2880 & 2014-04-29 \\   
PTF10qaf    & II  &  [1] & 0.284 & 23:35:42.89 & $+$10:46:32.9 & IMACS & 3.4 & 3600 & 2013-09-27 \\     
PTF10vqv  & I  &  [8]  & 0.452 & 03:03:06.84 & $-$01:32:34.9 & X-shooter & 0.7 & 3060 & 2013-09-10 \\  
SN 2011ke$^2$  & I  &  [9] & 0.143 & 13:50:57.77 & $+$26:16:42.8 & FORS2 & 11.3 & 1800  & 2013-05-30 \\           
SN 2011kf$^3$  & I  &  [9]  & 0.245 & 14:36:57.53 & $+$16:30:56.6 & FORS2 & 11.4 & 4500  & 2013-05-30 \\       
PTF11dsf  &II  &  [10]  & 0.385 & 16:11:33.55 & $+$40:18:03.5 & OSIRIS & 12.1 & 2700 & 2013-05-05$^{*}$ \\  
SN 2012il$^4$  &I  &  [9]   & 0.175 & 09:46:12.91 & $+$19:50:28.7 & IMACS & 3.4 & 4320  & 2013-02-08\\         
SSS120810  & I  &  [11] & 0.156 & 23:18:01.82 & $-$56:09:25.7 & FORS2 & 11.3 &  3600 & 2013-05-30 \\   
PTF12dam & R  &  [12]   & 0.107  & 14:24:46.20 & $+$46:13:48.3 & OSIRIS & 3.6 & 2400$^c$ & 2014-02-28 \\  
\hline
SNLS 06D4eu  & I  &  [13]  & 1.588 & 22:15:54.29 & $-$18:10:45.6 & X-shooter & 2.2 & 7200 & 2010-07-09 \\  
SN 2007bi  & R  &  [14,15]  & 0.128 & 13:19:20.00 & $+$08:55:44.0 & FORS2 & 11.3 & 14400 & 2008-04-10$^{**}$ \\  
SN 2008am  & II  &  [16]  & 0.233 & 12:28:36.25 & $+$15:35:49.1 & LRIS & 6.0  & 2200 & 2010-01-09\\
SN 2010gx$^5$  & I  &  [4,17]  & 0.230 & 11:25:46.71 & $-$08:49:41.4 & GMOS-N &  13.4  & 5200 & 2011-12-23$^{***}$\\      
PTF10hgi  & I  &  [9]  & 0.099 & 16:37:47.04 & $+$06:12:32.3 & FORS2 & 11.7 &  13600$^d$ & 2011-06-30$^{**}$ \\     
PS1-10bzj  & I  &  [18]  &  0.649 & 03:31:39.82 & $-$27:47:42.1 & GMOS-S & 9.1 & 2700 & 2011-04-03$^{*}$\\   
PS1-11ap & R  &  [19]  & 0.524  &  10:48:27.73 & $+$57:09:09.2  & GMOS-N & 12.5 & 7200 &  2011-12-27$^{**}$ \\    
\hline
\end{tabular}
\\ 
{\bf Alternative SN names}:  $^1$~PTF09cwl, $^2$~CSS110406:135058+261642, PTF11dij, PS1-11xk, $^3$~CSS111230:143658+163057, $^4$~CSS120121:094613+195028, PS1-12fo, $^5$~CSS100313:112547-084941, PTF10cwr. \\
{\bf References}: 
[1]  \cite{2012Sci...337..927G}, [2] \cite{2012A&A...541A.129L}, [3] \cite{2008ApJ...686..467S}, [4]  \cite{2011Natur.474..487Q},  [5]  \cite{2012AAS...21943604V}, [6] \cite{2013MNRAS.431..912Q}, [7] \cite{2010ATel.2634....1Q}, [8]  \cite{2010ATel.2979....1Q}, [9] \cite{2013ApJ...770..128I},  [10] \cite{2011ATel.3465....1Q}, [11] \cite{2014MNRAS.444.2096N}, [12]  \cite{2013Natur.502..346N}, [13] \cite{2013ApJ...779...98H}, [14]   \cite{2009Natur.462..624G}, [15] \cite{2010A&A...512A..70Y}, [16] \cite{2011ApJ...729..143C}, [17]  \cite{2010ApJ...724L..16P}, [18] \cite{2013ApJ...771...97L}, [19] \cite{2014MNRAS.437..656M}. \\
{\bf Notes}:  The coordinates refer to the SN.
$^a$~OSIRIS on GTC, IMACS on Magellan, X-shooter and FORS2 on VLT, LRIS on Keck, GMOS-N and -S on Gemini-N and -S, respectively, 
$^b$~As measured directly on the strongest skylines between 5500$-$6300 \AA\ (except SNLS~06D4eu for which the resolution refers to the NIR arm),
$^c$~Shared between two different grisms,
$^d$~Data obtained under ESO programme 087.D-0601 (PI:~P.~Mazzali),
$^*$~Mild or possible SN contamination, 
$^{**}$~Significant SN contamination, 
$^{***}$~Supplemented in the blue by a SN spectrum from \cite{2010ApJ...724L..16P} \citep[see][]{2013ApJ...763L..28C}. 
\end{minipage}
\end{table*}

\begin{table*}
 \centering
 \begin{minipage}{152mm}
  \caption{Measured and derived properties of the SLSN host galaxies. The SNe are grouped by type and then listed by discovery year.
  For three SNe we provide measurements at two different locations.  For these cases we only quote the total stellar mass of the galaxy.}
    \label{tab:res}
  \begin{tabular}{@{}lccccccc@{}}
\hline
SLSN host & Type & $M_B$ & Stellar mass & $E(B-V)$  & SFR (H$\alpha$)  & $W_{\rm{H}\alpha}$    & $W_{\lambda5007}$      \\ 
                  &              & (AB mag) & ($\log$ $M_{\sun}$) & (mag)  & ($M_{\sun}$ yr$^{-1}$)  & (\AA)    & (\AA)      \\ 
\hline 
SN 1999as & R & -18.02 & 8.60$^{+0.12}_{-0.09}$  &  0.66 (0.46) &   0.18 (0.19) &  10.67 (0.84) &   4.41 (0.78) \\ 
 -- SN location &   & \ldots & \ldots &   0.25 (0.17) &   0.04 (0.02) & 179.78 (12.53) & 102.57 (7.37) \\ 
SN 2007bi & R & -16.69 & 7.54$^{+0.28}_{-0.28}$  &  0.13 (0.16) &   0.02 (0.01) &   $>$  13.42  &   $>$   7.33$^*$  \\ 
SN 2010kd & R & -15.80 & 7.85$^{+0.63}_{-0.43}$  &  0.06 (0.06) &   0.07 (0.01) & 233.96 (3.29) & 190.90 (2.34) \\ 
PS1-11ap & R & -19.32 & 8.63$^{+0.15}_{-0.17}$  & -0.55 (0.33) &   0.57 (0.45) & \ldots &   $>$  42.31$^*$  \\ 
PTF12dam & R & -19.14 & 8.24$^{+0.39}_{-0.08}$  &  0.03 (0.01) &   4.83 (0.09) & 731.14 (0.53) & 794.25 (0.64) \\ 
SN 2006oz & I & -16.16 & 8.54$^{+0.28}_{-0.19}$  &  0.18 (0.41) &   0.13 (0.12) &  52.38 (5.29) &  33.41 (3.84) \\ 
SNLS 06D4eu & I & \ldots & 9.33$^{+0.34}_{-0.35}$  & -0.22 (0.15) &   3.01 (1.04) &  77.53 (3.67) &  99.06 (4.81) \\ 
PTF09cnd & I & -17.70 & 8.12$^{+0.17}_{-0.18}$  &  0.18 (0.11) &   0.21 (0.05) &  61.60 (0.65) &  45.63 (2.30) \\ 
SN 2009jh & I & -15.77 & 7.61$^{+0.75}_{-0.47}$  &\ldots &   $<$   0.01  & \ldots & \ldots \\ 
SN 2010gx & I & -17.28 & 7.80$^{+0.08}_{-0.42}$  &  0.01 (0.15) &   0.46 (0.16) & 302.88 (2.10) & 329.74 (2.81) \\ 
PTF10hgi & I & -16.93 & 7.23$^{+0.10}_{-0.06}$  &  0.91 (0.42) &   0.04 (0.04) &   $>$  13.18  &   $>$   6.55$^*$  \\ 
PTF10vqv & I & -18.30 & 8.09$^{+0.37}_{-0.23}$  & -0.43 (0.01) &   0.45 (0.01) & 139.08 (24.47) & 524.23 (87.74) \\ 
PS1-10bzj & I & -18.11 & 8.61$^{+0.47}_{-0.25}$  &  0.19 (0.16) &   6.04 (2.28) & \ldots & 384.62 (4.21) \\ 
SN 2011ke$^{**}$  & I & -16.87 & 8.04$^{+0.16}_{-0.13}$  &  0.02 (0.02) &   0.44 (0.02) & 820.92 (6.28) & 760.79 (5.31) \\ 
 -- tadpole tail$^{**}$ &   & -16.97 & \ldots &   0.23 (0.07) &   0.03 (0.01) &  31.60 (0.99) &  21.13 (0.93) \\ 
SN 2011kf & I & -16.33 & 7.58$^{+0.20}_{-0.28}$  &  0.22 (0.13) &   0.15 (0.05) & 174.21 (8.19) & 119.35 (6.02) \\ 
SN 2012il & I & -17.69 & 7.96$^{+0.19}_{-0.20}$  &  0.04 (0.07) &   0.40 (0.07) & 182.40 (5.09) & 265.65 (4.36) \\ 
SSS120810 & I & -16.62 & 7.96$^{+0.24}_{-0.26}$  &  0.05 (0.24) &   0.06 (0.04) & 115.12 (7.09) &  33.91 (3.82) \\ 
SN 1999bd & II & -18.99 & 9.52$^{+0.26}_{-0.24}$  &  0.49 (0.13) &   1.09 (0.34) &  36.50 (1.03) &   9.11 (0.89) \\ 
SN 2006tf & II & -16.72 & 7.70$^{+0.04}_{-0.04}$  &  0.43 (0.12) &   0.09 (0.03) &  41.09 (1.15) &  16.45 (0.77) \\ 
SN 2008am & II & -20.12 & 9.13$^{+0.19}_{-0.14}$  &  0.35 (0.12) &   1.38 (0.39) &  21.55 (1.13) &  11.07 (0.75) \\ 
PTF10heh & II & -18.00 & 8.36$^{+0.18}_{-0.17}$  & -0.17 (0.11) &   0.15 (0.04) & 125.60 (4.80) &  77.01 (3.76) \\ 
PTF10qaf & II & -21.72 & 10.24$^{+0.22}_{-0.17}$  &  0.25 (0.12) &   3.13 (0.89) &  17.66 (0.40) &   1.86 (0.42) \\ 
 -- SN location &   & \ldots & \ldots &  -0.36 (0.44) &   0.18 (0.19) &  42.44 (4.65) &  19.66 (5.83) \\ 
PTF11dsf & II & -20.07$^{***}$ & 8.97$^{+0.11}_{-0.11}$  &  0.02 (0.06) &   2.79 (0.42) &  39.56 (1.16) & 123.37 (2.38) \\ 
\hline 
\end{tabular}
\\ 
$^*$~These equivalent widths were measured on spectra with strong SN contamination and they are therefore strict lower limits.\\
$^{**}$~SN~2011ke exploded at the `head' of a `tadpole' galaxy. The other extraction refers to the `tail' of this galaxy.\\
$^{***}$~This magnitude differs considerably from what is measured in SDSS. There is the possibility that our observations are still contaminated by the SN or by an AGN (see Appendix~\ref{app:11dsf}).
\end{minipage}
\end{table*}

Our present spectroscopic sample (16 objects) is described in Table~\ref{tab:log}, where the essential information on the SLSNe is given, including their sub-class.
We follow the nomenclature of \cite{2012Sci...337..927G}, i.e. we separate H-poor SLSNe into SLSN-I and SLSN-R types according to their light curve evolution. 
SLSNe-I decline with a rate faster than the radioactive decay of $^{56}$Co \citep{2011Natur.474..487Q}, while SLSNe-R demonstrate an enduring light curve \cite[e.g.][]{2009Natur.462..624G}. 
While \cite{2013Natur.502..346N} have shown that there is a link between these classes, their phenomenological difference might still reflect important physical differences.
As a starting point, in this paper we report their host properties separately (e.g. Table~\ref{tab:res}). With the present sample sizes, however, we are not able to find significant differences between their environments and in many cases we group them together and treat them collectively as H-poor SLSNe.
In contrast to H-poor events, SLSNe-II show hydrogen in their spectra. All SLSNe-II in this paper belong to the IIn class, i.e. they show narrow H emission lines.  
No rare IIL events are included \citep[such as SN~2008es; ][]{2009ApJ...690.1313G,2009ApJ...690.1303M}.

In addition, we complement our SUSHIES sample with all spectra that are publicly available in telescope archives.
Details of the archival spectra observations appear below the horizontal line in Table~\ref{tab:log} (seven hosts).
For consistency, all these spectra have been re-reduced\footnote{with the exception of the SN 2008am host spectrum, which was kindly provided to us by Manos Chatzopoulos.} and re-analysed by us. 
There are only five additional SLSN hosts with spectra in the literature \citep{2014ApJ...787..138L} that are not treated here. 
We do not study the host of SLSN-II CSS100217 \citep{2011ApJ...735..106D}, which is 
known to have a significant active galactic nucleus (AGN) contribution. However, we do include it in the discussion to emphasize the difference between the H-poor and H-rich SLSN environments.

\subsection{Reductions and analysis}

All spectra have been reduced, extracted, wavelength and flux-calibrated in a standard way with a combination of {\sc iraf}\footnote{{\sc iraf} is distributed by the National Optical Astronomy Observatory, which is operated by the Association of Universities for Research in Astronomy (AURA) under cooperative agreement with the National Science Foundation.} and custom routines. 
Subsequently, the spectra were scaled with the host photometry (presented in Schulze et al., in prep.) and corrected for foreground extinction \citep{2011ApJ...737..103S} based on a \cite{1989ApJ...345..245C} law, assuming $R_V = 3.1$.
In most cases, scaling with one photometric band proved satisfactory (demonstrating that the flux calibration of the spectra is generally accurate), but in a handful of cases, we had to warp the spectra in order to match multi-band photometry. 
A few spectra, especially archival and for H-poor SLSNe (Table~\ref{tab:log}), suffer from significant SN contamination, as the main objective of these observations was to study the SN.
These spectra were scaled with the appropriate SN photometry reported in the corresponding references (Table~\ref{tab:log}).
The final host spectra are shown in Figs.~\ref{fig:spec1}--\ref{fig:spec4}.
We do not show spectra containing the SNe and that have been shown elsewhere.

Line fluxes and equivalent widths (EWs) were measured by fitting Gaussian profiles to the emission lines. 
Weaker lines (such as H$\gamma$, [\mbox{O\,{\sc iii}}] $\lambda$4363 and [\mbox{N\,{\sc ii}}] $\lambda$6584) were fit together with stronger lines (mainly [\mbox{O\,{\sc iii}}] $\lambda$5007 and H$\alpha$), in order to keep their central wavelength and FWHM constrained.
We only found gaussians to be a bad approximation for the line profiles of one galaxy in which case direct integration was used. 
When either [\mbox{O\,{\sc iii}}] $\lambda$4959 or $\lambda$5007 were obviously affected by skylines or telluric absorption and their measured ratio deviated significantly  from the canonical 1:3 ratio, we used this ratio and the line that was not affected in order to calculate the flux of the affected line.
For spectra contaminated by SN continuum, the measured EWs are  strict lower limits to the true EW of the lines.
The fluxes of the nebular lines are not affected by the presence of the SN. 
The measured line fluxes for the most important lines are reported in Table~\ref{tab:fluxes}. 
If a line was not detected, we provide 2$\sigma$ upper limits.
In the case of SN~2009jh we detected neither continuum nor line emission from the host galaxy and we provide upper limits for [\mbox{O\,{\sc iii}}] $\lambda$5007 and H$\alpha$, which are expected to give the most constraining non-detections. 
EWs for [\mbox{O\,{\sc iii}}] $\lambda$5007 ($W_{\lambda5007}$) and H$\alpha$  ($W_{\rm{H}\alpha}$) are given in Table~\ref{tab:res}. All EWs quoted in this paper are reported in the rest frame (unless otherwise indicated). 
In a handful of cases where the host was spatially resolved or displayed multiple components it was possible to extract spectra at different locations, including the explosion location. 
The resulting measurements for these cases are reported separately in Tables~\ref{tab:res}, \ref{tab:res2}  and~\ref{tab:fluxes}.

The host galaxy extinction has been estimated from the Balmer decrement (from the ratio H$\alpha$/H$\beta$ or H$\gamma$/H$\beta$ when H$\alpha$ was not available) and the host galaxy spectrum was corrected accordingly, assuming case B recombination \citep{1989agna.book.....O}. In a few cases $E(B-V)$ was found to be (marginally) negative so no correction was applied.

\begin{table*}
 \centering
 \begin{minipage}{135mm}
  \caption{Derived metallicities for the SLSN hosts in different scales. The units are always 12 + log(O/H).  The quoted errors are only the measurement errors. Systematic errors for the individual metallicity scales are not included. In case it is not possible to select between two metallicity branches, both solutions are shown. The last column shows the ionization parameter in units of cm~s$^{-1}$. The SNe are grouped by type and then listed by discovery year. For three SNe we provide measurements at two different locations.}
    \label{tab:res2}
  \begin{tabular}{@{}lccccccc@{}}
\hline
SLSN host  & Type  & M91 & N2 & O3N2 & KK04   & direct ($T_e$)   & $\log{(q)}$    \\ 
\hline 
SN 1999as & R & \ldots &   8.56 (0.06) &   8.39 (0.07) & \ldots & \ldots & \ldots \\ 
 -- SN location &   & \ldots &   $<$   8.29  &   $<$   8.22  & \ldots & \ldots & \ldots \\ 
SN 2007bi & R &   8.20 (0.12) &   8.20 (0.06) &   8.19 (0.05) &   8.37  & \ldots &   7.53  \\ 
SN 2010kd & R &   7.93 (0.06) &   8.07 (0.05) &   8.02 (0.04) &   8.17  &   7.50 (0.06) &   8.03  \\ 
PS1-11ap & R &  8.16/ 8.49 & \ldots & \ldots &  8.34/ 8.57 & \ldots &  7.63/ 7.74 \\ 
PTF12dam & R &   8.17 (0.01) &   8.09 (0.01) &   8.01 (0.01) &   8.37  &   8.06 (0.04) &   8.05  \\ 
SN 2006oz & I &   8.36 (0.27) &   $<$   8.70  &   $<$   8.41  &   8.41  & \ldots &   7.54  \\ 
SNLS 06D4eu & I &   8.54 (0.08) &   8.28 (0.04) &   8.23 (0.03) &   8.64  & \ldots &   7.90  \\ 
PTF09cnd & I &  8.33/ 8.38 &   8.24 (0.06) &   8.22 (0.05) &  8.37/ 8.52 & \ldots &  7.52/ 7.46 \\ 
SN 2009jh & I & \ldots & \ldots & \ldots & \ldots & \ldots & \ldots \\ 
SN 2010gx & I &   7.98 (0.04) &   7.92 (0.04) &   7.93 (0.02) &   8.20  &   7.36 (0.04) &   8.01  \\ 
PTF10hgi & I & \ldots &   8.38 (0.05) &   8.26 (0.06) & \ldots & \ldots & \ldots \\ 
PTF10vqv & I &   7.98 (0.01) &   $<$   8.12  &   $<$   8.07  &   8.21  &   7.77 (0.20) &   7.97  \\ 
PS1-10bzj & I &   8.09 (0.06) & \ldots & \ldots &   8.30  &   8.13 (0.04) &   8.14  \\ 
SN 2011ke & I &   7.83 (0.02) &   $<$   7.83  &   $<$   7.91  &   8.09  &   7.55 (0.04) &   7.97  \\ 
 -- tadpole tail &   &  8.20/ 8.53 &   $<$   8.29  &   $<$   8.25  &  8.19/ 8.65 & \ldots &  7.50/ 7.34 \\ 
SN 2011kf & I &   8.14 (0.12) &   $<$   8.29  &   $<$   8.21  &   8.33  & \ldots &   7.80  \\ 
SN 2012il & I &   8.10 (0.06) &   $<$   8.10  &   $<$   8.03  &   8.31  &   7.81 (0.07) &   8.08  \\ 
SSS120810 & I &   7.97 (0.19) &   $<$   8.23  &   $<$   8.27  &   8.18  & \ldots &   7.44  \\ 
SN 1999bd & II &   8.44 (0.09) &   8.62 (0.02) &   8.52 (0.02) &   8.58  & \ldots &   7.24  \\ 
SN 2006tf & II &  8.27/ 8.46 &   8.27 (0.03) &   8.27 (0.03) &  8.32/ 8.58 & \ldots &  7.40/ 7.31 \\ 
SN 2008am & II &   8.48 (0.07) &   8.35 (0.03) &   8.35 (0.02) &   8.61  & \ldots &   7.46  \\ 
PTF10heh & II &   8.67 (0.05) &   8.33 (0.03) &   8.31 (0.02) &   8.81  & \ldots &   7.97  \\ 
PTF10qaf & II &   8.85 (0.04) &   8.74 (0.02) &   8.68 (0.04) &   9.01  & \ldots &   7.48  \\ 
 -- SN location &   &   8.68 (0.20) &   8.48 (0.12) &   8.44 (0.09) &   8.84  & \ldots &   7.69  \\ 
PTF11dsf & II &  8.12/ 8.51 &   $<$   8.18  &   $<$   8.15  &  8.31/ 8.60 & \ldots &  7.66/ 7.80 \\ 
\hline
\end{tabular}
\end{minipage}
\end{table*}

We used different strong-line diagnostics to compute the host metallicity in different scales.
A direct comparison between metallicities is only meaningful when reported in a common scale as large systematic differences between different calibrations exist \citep[][and references therein]{2008ApJ...681.1183K}. 
In  Table~\ref{tab:res2} we provide metallicities in  the following scales: M91  \citep{1991ApJ...380..140M}, KK04 \citep{2004ApJ...617..240K}, N2 and O3N2 \citep{2004MNRAS.348L..59P}.
In many cases, predominantly for galaxies hosting H-poor SLSNe, the [\mbox{N\,{\sc ii}}] line was not detected (an indication of low metallicity) and only an upper limit for the N2 and O3N2 scales can be computed. For the R23-based scales (M91 and KK04) that are double-valued, whenever we are not able to break the degeneracy \citep[using criteria in][]{2006A&A...459...85N,2008ApJ...681.1183K}, we report the values for both branches. In addition, we quote the ionization parameter $\log{(q)}$ that determines how strongly ionized the gas is, as computed iteratively together with the KK04 metallicity \citep{2008ApJ...681.1183K}. For a sub-sample of seven SLSN hosts (all H-poor) we detect the auroral [\mbox{O\,{\sc iii}}] $\lambda$4363 line and we are able to compute the metallicity through the direct ($T_e$) method. 

Finally, we estimate the star formation rate (SFR) based on the H$\alpha$ (or scaled H$\beta$) flux and the relation in \cite{1998ARA&A..36..189K}, assuming an initial mass function (IMF) from \cite{2003PASP..115..763C}.

In addition to the spectroscopic properties of the SLSN hosts, for the discussion presented in this paper, we require the galaxy stellar mass $M_{*}$.
The masses are obtained through SED fitting and they are reported in Table~\ref{tab:res}. 
The SED fitting is done by using {\it LePhare} \citep{1999MNRAS.310..540A,2006A&A...457..841I} in a way similar to \cite{2011A&A...534A.108K}.
More details will be given in Schulze et al. (in prep.) where the SED fits will also be presented.
Briefly, we used a grid of $\sim 10^{6}$ synthetic galaxy templates based on \citet{2003MNRAS.344.1000B} with different star formation histories (SFH), ages, metallicities and dust reddenings assuming a \citet{2003PASP..115..763C} IMF to fit the available photometry. The attenuation of star light by dust follows \citet{2000ApJ...533..682C}, and the contribution of emission lines to the broad-band magnitudes is taken into account through their relation to the galaxy's SFR \citep{1998ARA&A..36..189K}. The stellar mass that we provide is the median of the probability-weighted distribution of the parameter, with errors denoting the 1$\sigma$ bounds on the distribution.
Finally, in Table~\ref{tab:res}, we report the galaxies' absolute magnitude $M_B$. In most cases this has been computed by using a full $K$-correction  \citep{2002AJ....124..646H} based on the photometry and the galaxy spectrum itself. For spectra where the continuum signal was too weak, we used a simpler $2.5 \log{(1+z)}$ correction, accounting for cosmological expansion.

\subsection{Other galaxy samples}
\label{subsec:compsampl}

\subsubsection{GRB hosts}

To form a meaningful comparison sample for our SUSHIES spectroscopic sample we have selected all (long) GRB hosts with emission-line measurements reported in the literature.
To this end, we have extensively used the  GHostS data base\footnote{\url{www.grbhosts.org}} \cite[][]{2009ApJ...691..182S}.
Our final GRB sample consists of 23 galaxies \citep{2010A&A...514A..24H, 2010AJ....139..694L, 2005A&A...444..425R, 2006Natur.444.1050D, 2007A&A...464..529W, christensen08, 2008ApJ...676.1151T, Lev020819, 2011A&A...535A.127V, 2011AN....332..283P, 2012A&A...546A...8K, 2012PASJ...64..115N, 2014A&A...566A.102S}, 
of which a handful also have resolved spectroscopy at the site of the explosion.
We have computed line flux ratios and have determined host extinction, metallicities and SFRs in the same way as for SLSNe and we have used the stellar masses that are quoted in GHostS.
We have checked that the GHostS stellar masses agree well (within 1$\sigma$) with the ones produced by our SED fitting method for the same data \citep{2011A&A...534A.108K}.

All GRB hosts lie at $z<1$, with the exception of one galaxy at $z=1.6$ \citep{2012A&A...546A...8K}. 
Since this is also the case for the SUSHIES sample \citep[i.e. all hosts at $z<1$, except one at $z=1.6$; ][]{2013ApJ...779...98H}, we do include these two individual objects at higher redshift.
The median redshift for the literature GRB host sample is $z=0.47$, while the one for SLSN hosts is $z=0.31$. 
H-poor SLSNe have a median redshift of $z=0.34$ (with $z=0.39$ for SLSN-I and $z=0.20$ for SLSN-R), while SLSN-II are found at a median redshift $z=0.24$.
Typical dispersions (as measured by the standard deviation) are of the order of 0.2 in redshift. So although differences do exist, we demonstrate in  Section~\ref{subsec:bias} that the redshift ranges are similar enough to not cause large systematic effects. 

Since EWs are rarely reported in the literature, for the $W_{\lambda5007}$ distribution of GRB hosts we have used a different sample\footnote{unpublished data; Malesani et al., in prep.}, the one of all $z<1$ galaxies belonging to the TOUGH survey \citep{2012ApJ...756..187H}. This sample has the additional advantage that it is complete and unbiased, as it was selected independent of the GRB optical emission \citep{2012ApJ...756..187H}. The $z<1$ TOUGH sample consists of 12 hosts at a median redshift of $z=0.59$ (i.e. it is more distant on average than the literature sample).

\subsubsection{Extreme emission line galaxies}

As SLSN hosts often exhibit strong nebular emission, we wanted to compare our sample with other galaxies that have similar properties. To this end, we have chosen two samples of EELGs recently presented by \cite{2014arXiv1403.3441A,2014A&A...568L...8A}.
These are galaxies that were selected to have $W_{\lambda5007} > 100$ \AA\ in deep spectroscopic surveys, and specifically the zCOSMOS-20k \citep{2009ApJS..184..218L} and the VIMOS Ultra-Deep Survey \citep[VUDS;][]{2014arXiv1403.3938L}. 
These samples lie also at $z<1$ (median $z=0.48$ for 165 zCOSMOS and $z=0.57$ for 31 VUDS galaxies). 
EELGs are similar in properties to the more nearby blue compact dwarfs \citep[BCDs;][]{1981ApJ...247..823T},  \mbox{H\,{\sc ii}} galaxies \citep{1991A&AS...91..285T} and the SDSS colour-selected Green Peas  \citep{2009MNRAS.399.1191C, 2011ApJ...728..161I, 2010ApJ...715L.128A, 2012ApJ...749..185A}.
However, they offer the advantage that they consist of a large homogeneous sample  and that their redshift range is more similar to the SLSN and GRB hosts. 
For this reason, we preferred these as a comparison sample over the other starburst dwarfs.
Especially the zCOSMOS EELG sample is well-defined in the sense that it represents the tip of the $W_{\lambda5007}$ distribution of a much larger ($\sim$20,000) galaxy sample, which is purely magnitude limited ($I_{\rm{AB}} < 22.5$ mag) but with no further selection biases. 
All EELG spectroscopic properties were re-computed consistently with SLSN hosts, based on the measured emission line fluxes in \cite{2014arXiv1403.3441A,2014A&A...568L...8A}.

\subsubsection{Core-collapse SN hosts}

There is no suitable sample of regular core-collapse (CC) SN hosts in the literature that we can use to compare with our spectroscopic SUSHIES sample. 
\cite{2014ApJ...787..138L} use the GOODS sample \citep{F06, 2010MNRAS.405...57S}.
However, this sample is not good for our purposes as it does not contain spectroscopic properties (such as line fluxes and metallicities) 
but only properties derived by the photometry and SED fits. 
For this reason, and despite the redshift discrepancy, we have decided to use more nearby CC SN samples to overcome these problems. 
 \cite{2011A&A...530A..95L} and \cite{2012ApJ...758..132S} provide line fluxes for the hosts of Stripped Envelope (SE) CC SNe (i.e. CC SNe that have been stripped of their H).
We have used these data to determine line ratios, metallicities and ionization fractions, selecting only SNe that were discovered in impartial surveys (i.e. surveys that do not target specific galaxies).
EWs were only measured in the spectra of \cite{2011A&A...530A..95L} to which we have access.
In addition, we calculated the stellar masses for the sub-sample of these galaxies for which data is available in SDSS. 
This was done by running our SED fitter on the Petrosian SDSS $ugriz$ magnitudes in the same way that this was done for SLSNe.
A visual inspection was performed to make sure that the correct host was picked and that the correct magnitudes were used.
The final number of galaxies for which both stellar masses and (at least some) line fluxes are available is 38 (after rejecting a couple of ambiguous cases).
Due to various uncertainties affecting the absolute flux scaling (such as the usually unknown percentage of galaxy coverage, slit losses, and observational details) 
it is not possible to use the line fluxes of this spectroscopic sample in order to determine absolute quantities, such as SFR.
To compare the sSFR (derived from H$\alpha$) in CC SN hosts with the other galaxy samples in this paper, we have used a sample from  \cite{2012ApJ...759..107K} that contains this information, together with the galaxy stellar mass. 
We caution, however, that the SFRs quoted by  \cite{2012ApJ...759..107K} are those within the SDSS fibre (3\arcsec arcsec aperture) and thus represent a lower limit to the true values.
However, this effect should be less important for the hosts of impartially selected SNe that we use in this study, because they are typically smaller than the hosts of SNe found in targeted surveys \citep{2012ApJ...759..107K}.
The median redshift of all CC SN host samples used in this paper is $z\sim 0.04$ (see Section~\ref{subsec:bias} for a discussion on the effect of this potential bias).

\section{Results}
\label{sec:results}

\subsection{Mass and metallicity}

\begin{figure*}
\includegraphics[width=\textwidth]{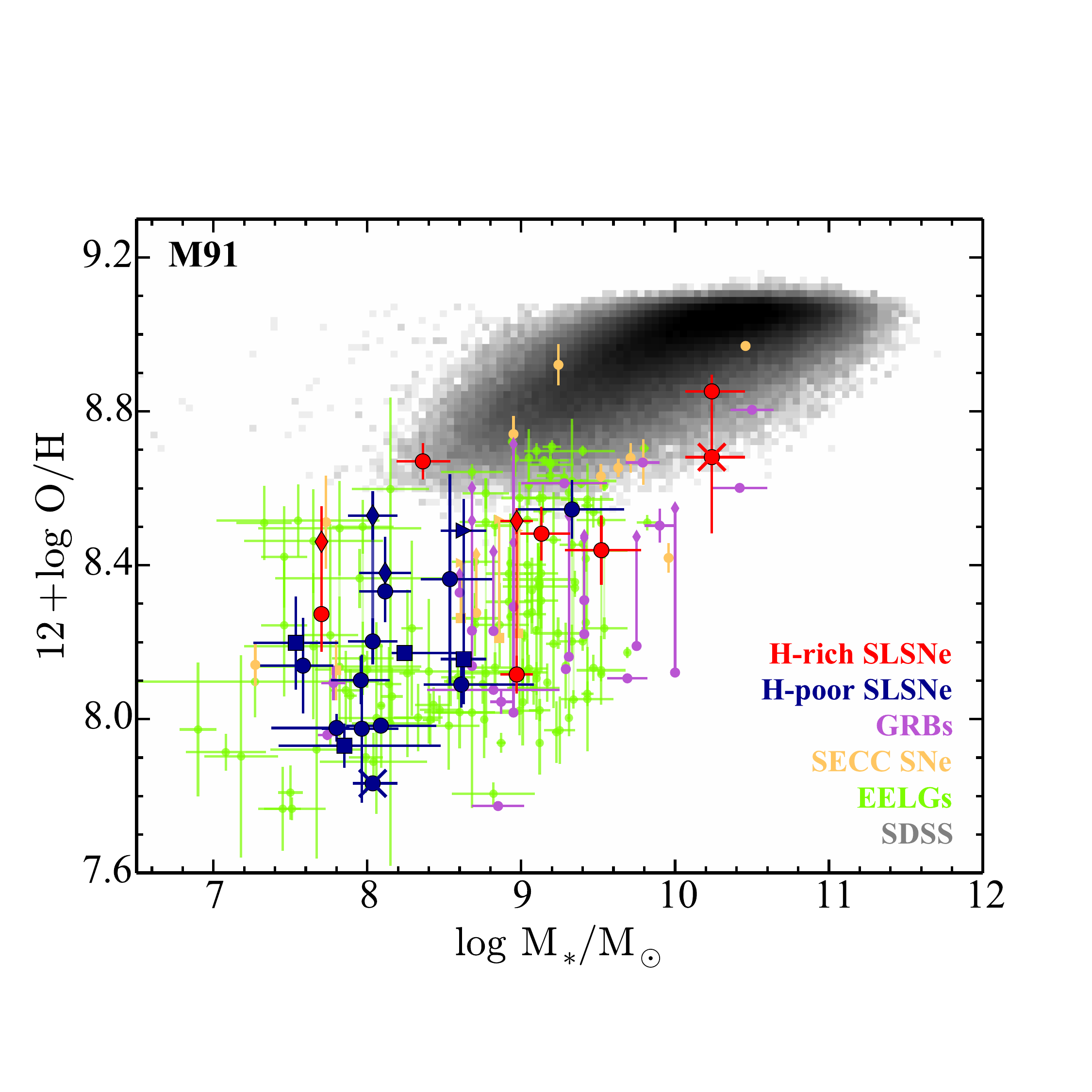}
\caption{
A mass-metallicity diagram for SLSN hosts in the R23  calibration of \protect\cite{1991ApJ...380..140M}. 
H-poor SLSN hosts are marked with blue symbols: circles for SLSNe-I and squares for SLSNe-R. 
H-rich SLSN hosts are marked in red. 
An overlaid X signifies a metallicity measurement at the exact explosion site.
Also shown are GRB hosts (violet), stripped envelope CC SN hosts from un-targeted surveys (orange circles for regular events and orange squares for broad-lined events), EELGs (green) and SDSS galaxies (gray). 
When it is not possible to distinguish between the 2 metallicity branches, both solutions are plotted and are connected with a solid line. 
In addition, for clarity, upper branch solutions are marked with different symbols than the lower branch: diamonds instead of circles and rightward pointing triangles instead of squares (applies to SLSNe-R and SNe~Ic-BL).
The metallicity error bars shown include only the flux measurement errors (not available for many hosts in the GRB literature sample).
The systematic uncertainty related to the M91 scale is $\sim0.15$ dex \protect\citep{2008ApJ...681.1183K}. Metallicities for all samples have been computed in a consistent manner based on measured emission line fluxes.
}
\label{fig:MZ}
\end{figure*}

A mass-metallicity graph \citep[e.g.][]{2004ApJ...613..898T} for SLSN and GRB hosts is presented in Fig.~\ref{fig:MZ}. 
For this graph we have selected the R23 scale, and specifically the M91 calibration, in order to maximize the number of both SLSN and GRB host metallicities. 
In addition, we show data from SDSS, based on the line fluxes provided by the MPA-JHU group \citep[e.g.][]{2003MNRAS.341...33K, 2004MNRAS.351.1151B}, the EELGs from \cite{2014arXiv1403.3441A,2014A&A...568L...8A} and a sample of stripped CC SN hosts discovered in galaxy impartial surveys \citep{2011A&A...530A..95L,2012ApJ...758..132S}.

A simple inspection of the graph  immediately shows that SLSN and GRB hosts do not occupy the same region as SDSS galaxies, which are more massive and more metal-rich. In contrast, they seem to occupy a region similar to the EELGs. This is unlike the SNe Ibc hosts which are well represented in the SDSS cloud. The issue of whether GRB hosts lie below the main mass-metallicity relation of star forming galaxies has been extensively discussed in the literature \cite[e.g.][]{2010AJ....139..694L,2011MNRAS.414.1263M,2013ApJ...774..119G}. An offset of 0.3~dex has also been found between SDSS galaxies and Green Peas of similar masses   \citep{2010ApJ...715L.128A}.
It is beyond the scope of this paper to determine whether SLSNe occur below the SDSS mass-metallicity relation.
It is possible that they also form an extension towards lower masses, as there is only a  small number of low-mass ($M <$ 10$^{8.5}$ $M_{\sun}$) galaxies in SDSS.

The derived host galaxy masses (Table~\ref{tab:res}) suggest that H-poor SLSNe hosts appear to be less massive than the GRB hosts: the majority of the former lie below $\log{M_{*}} =8.6$, while the opposite is true for the latter (Fig.~\ref{fig:MZ}). The formal $p$ value for a Kolmogorov-Smirnov (KS) test among the mass distributions presented in this paper is $1.7\times10^{-5}$. A statistically significant difference ($>2.7 \sigma$) between the mass distributions of H-poor SLSN and GRB hosts was also observed by \cite{2014ApJ...787..138L}, despite their conclusion that GRB and H-poor SLSN hosts are similar. We caution, however, that the galaxies in the SUSHIES spectroscopic sample are not suitable for studying mass distributions as they have been selected from galaxies for which spectroscopy is available. 
In any case, including fainter galaxies for which spectroscopy was not attempted will skew the mass distribution of H-poor SLSN hosts to even lower values with respect to GRB hosts.
We do not see any significant difference between the masses of SLSN-R and SLSN-I hosts ($p = 0.68$), while SLSNe-II are found in more massive hosts on average: a KS test gives $p = 0.04$ indicating that the difference with H-poor SLSN hosts  is statistically significant at the 2$\sigma$ level.

\begin{figure*}
\includegraphics[width=\textwidth]{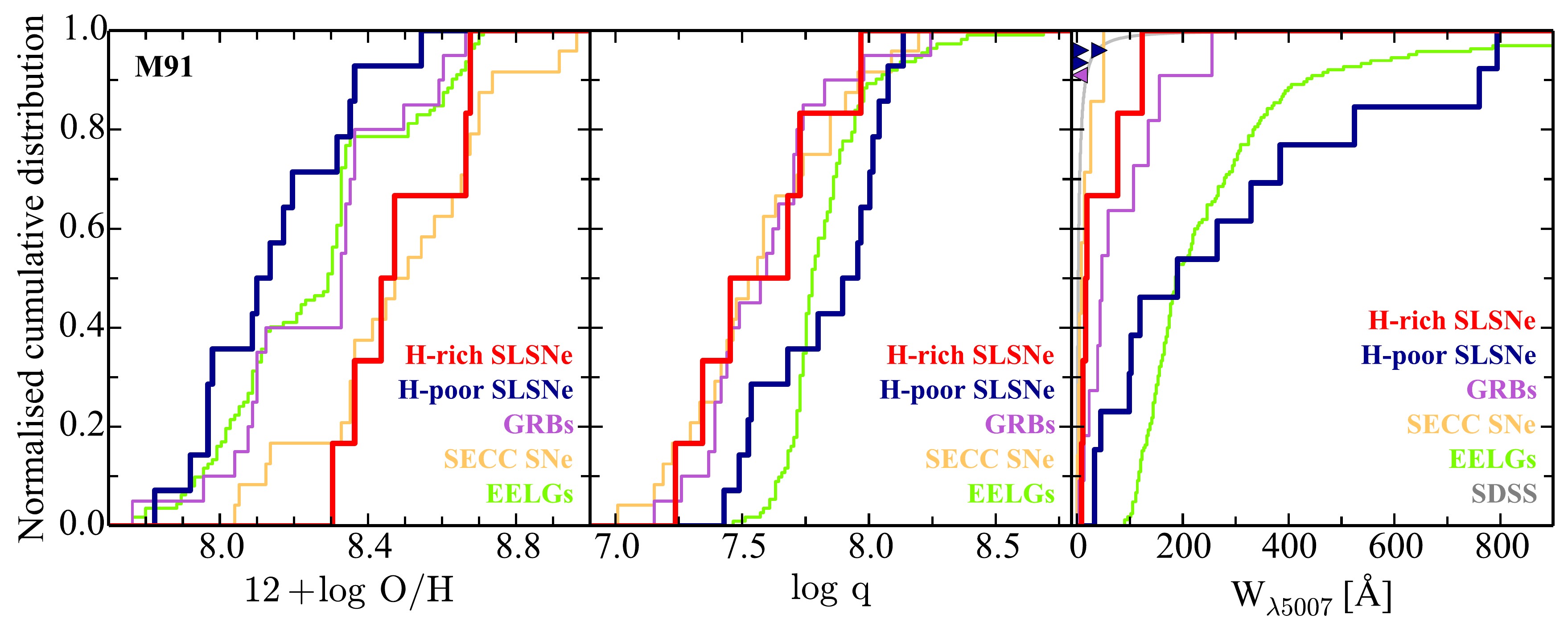}
\caption{Normalized cumulative distributions for the nebular properties of different galaxy samples. 
\textbf{Left:}  Metallicity (M91 scale). H-poor SLSNe are found in more metal-poor environments than H-rich events ($p = 0.006$). The metallicity distribution of H-poor SLSN hosts is statistically indistinguishable from those of GRB hosts and EELGs. 
\textbf{Middle:} Ionization parameter $\log(q)$. H-poor SLSNe are found in harder radiation fields than GRBs, H-rich SLSNe and stripped CC SNe.   
\textbf{Right}: The rest-frame EW of [\mbox{O\,{\sc iii}}] $\lambda$5007.  
Upper and lower limits are shown with triangles pointing left and right respectively.
H-poor SLSNe occur in galaxies with stronger emission lines than CC~SNe, H-rich SLSNe and GRBs. 
The EELGs \protect\citep{2014arXiv1403.3441A} were  by definition selected to have $W_{\lambda5007} > 100$ \AA.
The GRB host sample in this panel (the $z<1$ TOUGH sample) is different than the GRB literature sample in the other figures.}
\label{fig:CDFs}
\end{figure*}

Figure~\ref{fig:CDFs} (left panel) shows the metallicity distributions for the galaxy samples shown in Fig.~\ref{fig:MZ}. 
Here, double-valued metallicities have been averaged for the galaxies that it was not possible to distinguish between the upper and the lower M91 branch (Table~\ref{tab:res2}).
The distributions of H-poor SLSN hosts, GRB hosts and EELGs are statistically indistinguishable, something that is also confirmed by KS tests ($p$ values $> 0.10$).
In general, H-poor SLSNe seem to occur in low-metallicity galaxies. The median in the M91 scale displayed is 0.27 $Z_{\sun}$ ($\log{(12+ \rm{O}/\rm{H})} = 8.12$) and the full range is 0.14--0.71 $Z_{\sun}$  \citep[where we are using the solar abundances from][]{2009ARA&A..47..481A}.
In contrast, H-rich SLSNe appear in more metal-rich galaxies (0.4 -- 1.0 $Z_{\sun}$) and the difference with H-poor SLSN hosts is significant at the 3$\sigma$ level ($p = 0.006$).
The distribution of H-rich SLSN host metallicities is instead consistent with the metallicity distribution of the (stripped) CC SNe ($p = 0.96$).

We conclude that metallicity is not an important parameter differentiating H-poor SLSN from  GRB environments \citep[see also][]{2014ApJ...787..138L}. It is, however, an important difference between H-poor and H-rich SLSN hosts.

\subsubsection{Direct metallicities}
\label{subsec:dirmet}

For seven H-poor SLSN hosts we were able to measure a more accurate metallicity based on the detection of [\mbox{O\,{\sc iii}}] $\lambda$4363 and the direct ($T_e$) method.
We use the nebular package in {\sc iraf}  \citep{1995PASP..107..896S} to determine the gas phase temperatures and densities from the measured emission line fluxes. To derive the total oxygen abundances we add the abundances from single and double ionized oxygen levels using the equations in \cite{2006A&A...448..955I}.
We thus increased the number of SLSN hosts \citep{2013ApJ...763L..28C,2013ApJ...771...97L,2014ApJ...787..138L} with a direct metallicity  measurement by three galaxies.
Our measurements suggest lower metallicities (0.04--0.27 $Z_{\sun}$) than those obtained using any other metallicity scale.
This is  because strong-line methods are not well calibrated at such low metallicities. The values obtained are similar to those of many EELGs with a detection of [\mbox{O\,{\sc iii}}] $\lambda$4363 \citep{2014arXiv1403.3441A}. Among these seven galaxies there are three where the metallicity is less than 0.07 $Z_{\sun}$, i.e. there is an unusually high percentage of extremely metal-poor galaxies among H-poor SLSN hosts. The corresponding percentage in SDSS is $<0.01\%$ \citep[][]{2008A&A...491..113P,2011ApJ...743...77M}. 


\subsection{Ionization and emission line strength}

In the middle panel of  Fig.~\ref{fig:CDFs} we show the ionization parameter for the same samples as in the left-hand panel. 
The ionization parameter $q$ is defined as the flux of the ionizing photons divided by the gas density.
We observe that H-poor SLSN hosts are found in environments where the gas is more ionized than stripped CC SNe, H-rich SLSNe and GRBs.
GRB hosts have a median $\log{(q)} = 7.59$, a value similar to those of SLSN-II and stripped CC SN hosts.
These three distributions are statistically indistinguishable with KS test $p$ values $>0.9$. 
In contrast, H-poor SLSN hosts have a median $\log{(q)} = 7.93$ and their  $\log{(q)}$ distribution is different than the one of  GRBs at a significance higher than 2$\sigma$ ($p = 0.023$). In addition, H-poor SLSNe are the only transients that are found in host galaxies that are not inconsistent with being drawn from the EELGs $\log{(q)}$ distribution  ($p = 0.08$).

What is especially striking in the case of H-poor SLSN hosts is the strength of the emission lines in their spectra (see Figs.~\ref{fig:spec1}--\ref{fig:spec4}).
The right-hand panel of Fig.~\ref{fig:CDFs} shows the distribution of EWs for the galaxy samples studied in this paper.
From all the galaxies that host different kinds of explosions, we see a sequence in the $W_{\lambda5007}$ distributions as we move from the environments of SNe~Ibc \citep{2011A&A...530A..95L} and SLSN-II (although these two first samples are small), to those of GRBs (the complete $z<1$ TOUGH sample) and finally to those of H-poor SLSNe. The latter have a median value  of 191~\AA\ and contain values reaching almost 800~\AA. 
We see no significant difference between SLSN-R and SLSN-I environments ($p=0.68$) and we therefore group them together.
For resolved galaxies, we have used the values at the SN explosion locations.
$W_{\lambda5007}$ lower limits (lines measured on spectra contaminated by SN continuum) appear as arrows in Fig.~\ref{fig:CDFs}.
The KS test $p$ values are 0.03 between H-poor and H-rich SLSN hosts (i.e. different at $>$ 2$\sigma$) and 0.11 between H-poor SLSN hosts and GRB hosts. 
In addition, the H-poor SLSN host distribution is the only distribution that is not inconsistent with the one of the zCOSMOS EELGs ($p = 0.07$, all others $<10^{-4}$), despite the fact that the latter were \textit{selected} to have $W_{\lambda5007} > 100$~\AA.

In our sample,  more than 50$\%$ of the H-poor SLSN hosts have $W_{\lambda5007} > 100$~\AA, which makes them \textit{de facto} EELGs. Although the exact number depends on how we treat resolved locations and lower limits, the high EELG representation within our H-poor SLSN host sample cannot have happened by chance. 
EELGs are a rare galaxy class \citep[less than 0.5\% in SDSS and $\sim$1\% in zCOSMOS;][]{2014arXiv1403.3441A} and 
a simple binomial test reveals that the chance of getting eight EELGs in a sample of 16 galaxies (as is the case for our H-poor SLSN hosts) is negligible ($p \sim$10$^{-12}$).
In fact, it would be highly improbable ($p = 0.05 \%$) to obtain even three galaxies with $W_{\lambda5007} > 100$~\AA\ by pure chance in such a small sample.

We conclude that H-poor SLSNe show a preference for exploding in EELGs, while this is not the case for H-rich SLSNe.
In addition, there is  evidence supporting that H-poor SLSN environments are more extreme  and have a harder ionization field 
than those of GRBs.  

\begin{figure*}
\includegraphics[width=\textwidth]{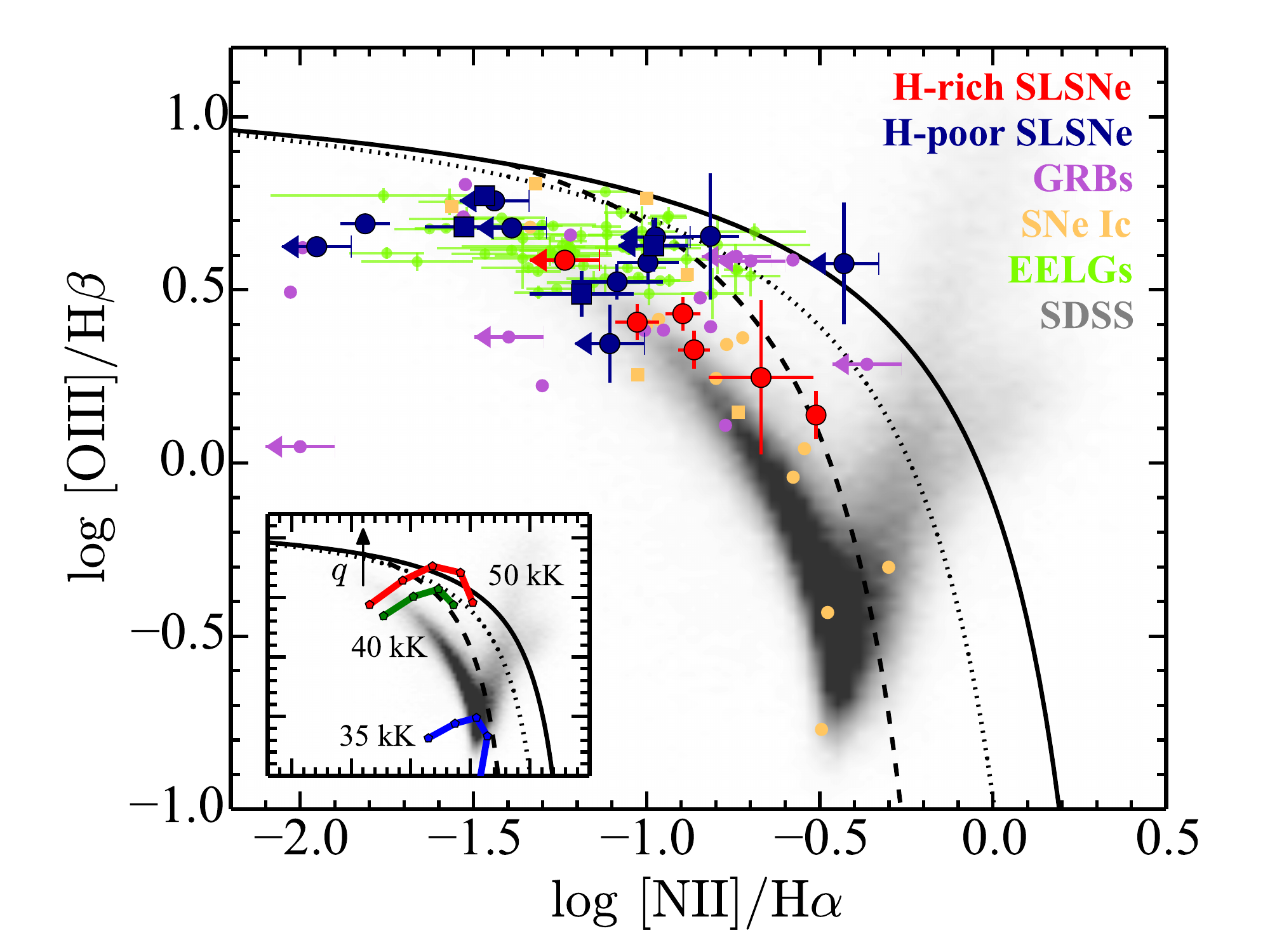}
\caption{BPT diagram for SLSN hosts and comparison galaxy samples. 
The symbols for SLSN-I, SLSN-R, SLSN-II, GRB hosts and EELGs are the same as in Fig.~\ref{fig:MZ}.
Upper limits (due to [\mbox{N\,{\sc ii}}] non-detections) are marked with arrows. 
SN~Ic hosts from impartial surveys are shown in orange.
Squares mark the environments of SNe~Ic-BL and circles those of regular SNe~Ic.
The solid and the dashed lines are the AGN separation lines from \protect\cite{2001ApJ...556..121K} and \protect\cite{2003MNRAS.341...33K}, respectively.
We also show an indicative separation line from the redshift-dependent models of \protect\cite{2013ApJ...774L..10K} at $z = 0.4$ (dotted line). 
The inset shows a grid of photoionization models from \protect\cite{2005MNRAS.361.1063P} that highlight the dependence on temperature and metallicity.
This grid has been calculated for an ionization parameter of $\log{(q)} = 7.7$ and for metallicities of 7.62, 7.92, 8.22, 8.62 and 8.92~dex (increasing from left to right on the isotemperature lines).  
 An arrow marks the qualitative dependence on the ionization parameter in a relevant location of the BPT diagram \protect\cite[e.g.][]{2013ApJ...774..100K}.
H-poor SLSNe, like EELGs, are found almost exclusively at locations with $\log([\mbox{O\,{\sc iii}}]/\rm{H}\beta) > 0.5$ for any given $[\mbox{N\,{\sc ii}}]/\rm{H}\alpha$ ratio. 
GRB hosts are also found in less ionized environments. 
Most H-rich SLSNe are consistent with the locus of SDSS star forming galaxies.}
\label{fig:BPT}
\end{figure*}

Another way to look at this is by examining a BPT \citep{1981PASP...93....5B} diagram (Fig.~\ref{fig:BPT}).  
We observe that  the H-poor SLSN location is similar to the one of  EELGs: at any given [\mbox{N\,{\sc ii}}]/H$\alpha$ ratio, the large majority of H-poor SLSN hosts have $\log($[\mbox{O\,{\sc iii}}]/H$\beta) > 0.5$, while this is only true for about half the GRB hosts. 
SLSN-II hosts seem to better follow the emission line ratios of the SDSS star forming galaxies. 
The same is true for impartially selected SNe~Ic \citep{2011A&A...530A..95L,2012ApJ...758..132S}, some of which are found in galaxies with very low $[\mbox{O\,{\sc iii}}]/$H$\beta$ ratios. 
We notice, however, that a significant fraction of \textit{broad-lined} SNe~Ic (Ic-BL) seem to explode in galaxies that have high $[\mbox{O\,{\sc iii}}]/$H$\beta$ ratios.
SNe~Ic-BL have been shown to be related to GRBs \citep[e.g.][]{hjorth2003} and have also spectroscopic similarities with SLSNe-I at post-maximum phases \citep{2010ApJ...724L..16P}.
Many studies suggest that these explosions occur in different environments than normal SNe~Ic \citep[e.g.][]{2010ApJ...721..777A,2011ApJ...731L...4M,2012ApJ...758..132S,2012ApJ...759..107K} and their locations on the BPT diagram might constitute another important difference. 
The location on the BPT diagram has a complex dependence on parameters like metallicity, ionization, age and temperature \cite[e.g.][]{2000ApJ...542..224D,2006ApJS..167..177D,2005MNRAS.361.1063P,2010AJ....139..712L,2014ApJ...795..165S,2015A&A...574A..47S}. In general, along the star forming sequence, there are locations of the graph that are strongly dependent on the ionization parameter \cite[e.g.][]{2010AJ....139..712L} and the most strongly ionized $\mbox{H\,{\sc ii}}$ regions are found on the upper left corner of the BPT diagram \citep{2015A&A...574A..47S}. In the region of interest for H-poor SLSN hosts, an increase in $q$ leads to an increase in the  $[\mbox{O\,{\sc iii}}]/$H$\beta$ ratio and we have marked this on our graph with a qualitative arrow similar to \cite{2013ApJ...774..100K}. But also the effective temperature can increase the $[\mbox{O\,{\sc iii}}]/$H$\beta$ for a given ionization parameter and metallicity \citep{2005MNRAS.361.1063P,2014ApJ...795..165S} by up to 0.15 dex (Fig.~\ref{fig:BPT}). This shows that the presence of very hot massive stars can be an important factor shaping the environments of H-poor SLSNe.

\subsection{Specific star formation rate}

\begin{figure*}
\includegraphics[width=\textwidth]{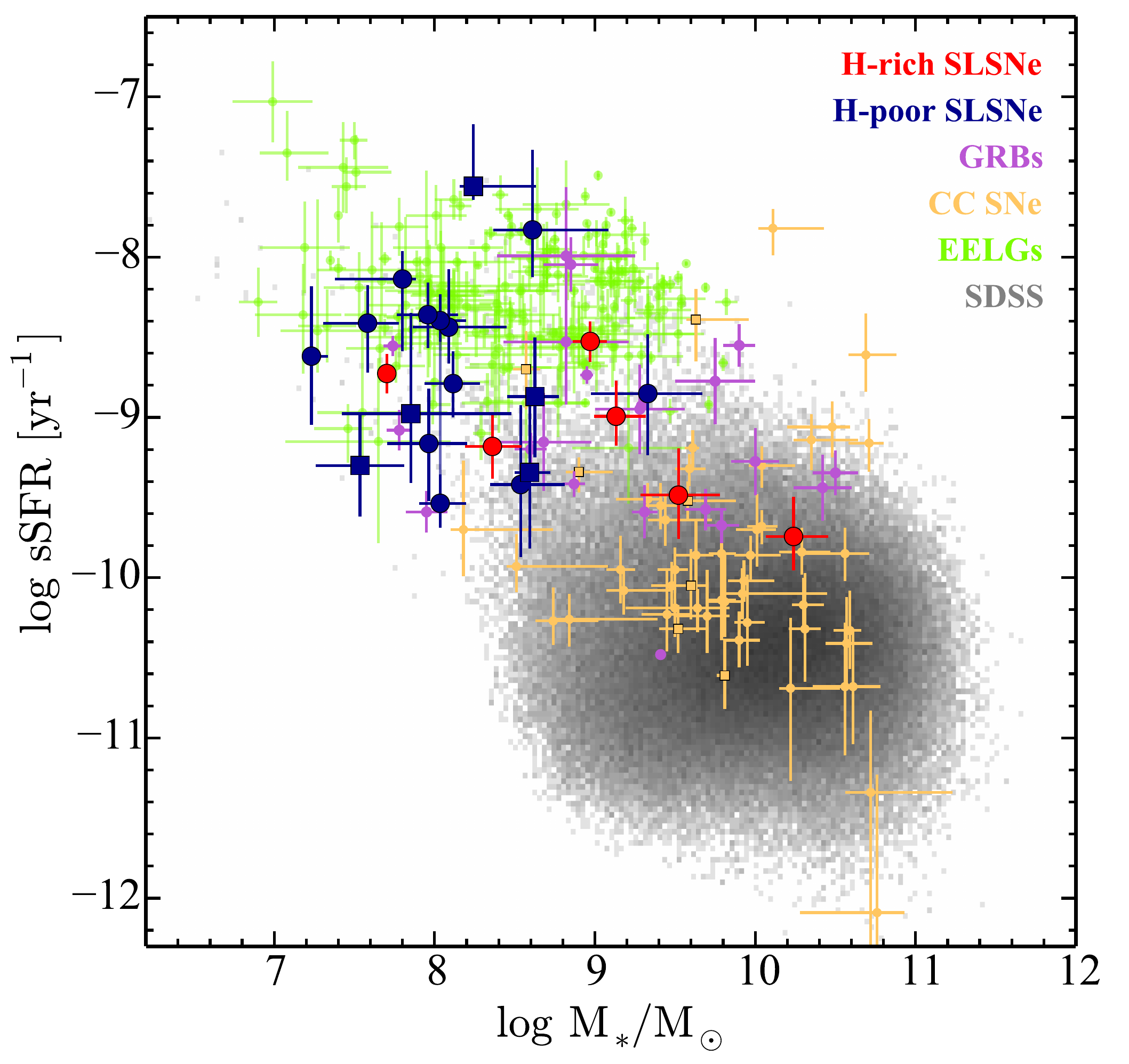}
\caption{Specific star formation rate versus stellar mass for various types of galaxies. The symbols are the same as in Fig.\ref{fig:MZ}.
The CC SN hosts are from the impartial sample of \protect\cite{2012ApJ...759..107K}. SNe~Ic are marked with a square instead of a circle.
The EELGs occupy a different location on this graph than the majority of SDSS galaxies.
H-poor SLSN hosts are broadly consistent with the locus of EELGs. H-rich SLSN and GRB hosts are also found in the SDSS locus (although on the upper end).
}
\label{fig:MsSFR}
\end{figure*}

EELGs are among the galaxies with the highest sSFR in the Universe \citep[e.g.][]{2009MNRAS.399.1191C,2014arXiv1403.3441A, 2014ApJ...791...17M}. 
In Fig.~\ref{fig:MsSFR}, we show that the location of EELGs  in the mass--sSFR plane is clearly separated from the main locus of SDSS galaxies by more than 2 dex in the vertical axis. Only few SDSS galaxies are found in the same region. Evolution with redshift cannot account for more than 0.5 dex of this difference (see Section~\ref{subsec:bias}).

SLSN hosts are also forming stars at a high pace relative to their small masses. 
Figure~\ref{fig:MsSFR} shows that H-poor SLSNe are mostly found in an area similar to the EELGs, although perhaps not at the most extreme region. GRBs and H-rich SLSN hosts are found both in the EELG and the SDSS locus. 
The median sSFR  (in $\log$) within this sample are $-8.53$ yr$^{-1}$ for SLSN-I hosts, $-8.98$ yr$^{-1}$ for SLSN-R hosts and $-9.09$ yr$^{-1}$ for SLSN-II hosts. 
The differences between the different SLSN subtypes are not statistically significant. 
In comparison, the median values obtained for the GRB literature spectroscopic sample and the EELG sample are  $-9.15$ yr$^{-1}$ and  $-8.21$ yr$^{-1}$, respectively. 
The $p$ value between the GRB and the H-poor SLSN hosts sSFR distributions is 0.25.
Standard deviations are of the order of 0.4--0.5 dex for these samples. 
In contrast, the impartially selected CC SNe of \cite{2012ApJ...759..107K}, are found in more quiescent galaxies, much more consistent with the general SDSS population ($\log{\rm{sSFR}} =  -10.0$, however one also needs to correct here for the flux lost outside the SDSS fibre).

\section{Discussion}
\label{sec:disc}

\subsection{A different progenitor channel for H-poor and H-rich SLSNe}
\label{sec:discHrich}

The first important result of this study is that  SLSNe-II come from different kinds of environments than SLSNe-I and SLSNe-R.
We have shown that they are found, on average, in galaxies that are more massive, more enriched in metals and where the gas is less ionized. 
Despite the small sample of SLSN-II, these differences appear to be statistically significant. 
This is a strong indication that if SLSNe-II constitute a uniform class, they represent a different phenomenon than H-poor SLSNe and that the progenitors of these events are different.
In the past, host galaxy differences have contributed to the early recognition that SNe~Ia constituted a distinct population from CC SNe and that short GRBs are due to a different physical mechanism than long GRBs.
All SLSNe-II studied here are scaled-up SNe-IIn and their extreme luminosity can be primarily attributed to circum-stellar material (CSM) interaction \citep[e.g.][]{2013MNRAS.428.1020M,2013ApJ...773...76C}.
CSM interaction is also one of the proposed luminosity sources for SLSNe-I \citep{2011ApJ...729L...6C}.
This possibility is not ruled out by the observed differences with SLSN-II environments.
The observed differences in the host environments suggest that, even if SLSNe-I are CSM powered, it is unlikely that SLSNe-II are simply SLSNe-I with some additional mass of H-rich CSM. 

Two more SLSNe-II (IIn) not covered in this study support the idea that H-rich SLSNe explode in different environments than H-poor SLSNe:  SN~2006gy \citep{2007ApJ...659L..13O,2007ApJ...666.1116S} and CSS100217 \citep{2011ApJ...735..106D}. 
The former was found in the massive NGC~1260 at super-solar metallicity, while the latter is a galaxy with a clear AGN contribution, i.e. galaxies  markedly different from those of H-poor SLSNe.
This supports that the differences seen within the SUSHIES spectroscopic sample  are unlikely to be due to the small size of SLSN-II sub-sample. 

Alternatively, SLSNe-II can represent a variety of events, as CSM interaction can in principle mask any kind of explosion below it \citep[e.g.][]{2015A&A...574A..61L}. In this case, a much larger sample of SLSNe-II will be needed in order to statistically uncover the potential differences between their environments. Furthermore, there are a few SLSN-II that do not show narrow H lines in their spectra  \citep{2009ApJ...690.1303M,2009ApJ...690.1313G,2014MNRAS.441..289B} and that have sometimes been referred to as Type~IIL. These objects are  not represented within the present SUSHIES spectroscopic sample and therefore our conclusions do not apply to those.

\subsection{A connection between H-poor SLSNe and EELGs}
\label{sec:discConn}

We have demonstrated that there is a preference for H-poor SLSNe to explode in EELGs and that this cannot be a chance coincidence. The implications of this finding are discussed below.

\subsubsection{Implications for the SN progenitors}
\label{subsec:ImplProg}

The extreme emission lines are thought to originate in starbursts at an early stage of their evolution \citep{2001A&A...374..800O,2011ApJ...728..161I,2013ApJ...766...91J,2014arXiv1403.3441A}. 
In addition, they appear in a transient phase in the life of a galaxy and this can be easily understood from the fact that such high sSFRs cannot be maintained for a long period. 
Dwarf galaxies have bursty  SFH, possibly interrupted by quiescent intervals, as can be seen in both observations \citep{2004MNRAS.348.1191T,2009ApJ...692.1305L} and simulations \citep{2012MNRAS.420.1294M,2014ApJ...792...99S}.
This can be explained by the fact that outflows, caused exactly by SN feedback, remove the gas from the dwarfs'  shallow potential well, thus quenching star formation \citep[e.g.][]{2010Natur.463..203G}, until this process starts over again when external or internal causes trigger a new wave of star formation.

The fact that H-poor SLSNe appear to be connected with this short-lived transient period suggests that they originate from the first stars in a newly born starburst. As the strength of the emission lines reduces with time \citep{1999ApJS..123....3L,2001A&A...375..814Z,2011MNRAS.415.2920I} and H-poor SLSNe appear in galaxies that show larger EWs than GRB hosts, we deduce that H-poor SLSNe occur earlier (on average) than GRBs in the life of a starburst.  
As a consequence, the progenitor stars of H-poor SLSNe must be very massive.
Indicatively, \cite{2008MNRAS.387.1227O} have estimated the stellar population age at the location of a GRB to be of the order of 5-8 Myr, implying a progenitor mass greater than 25 $M_{\sun}$ (or  50 $M_{\sun}$ depending on the age limit). Similar stellar population ages were found at the locations of other GRBs  \citep{2008ApJ...676.1151T,2010AJ....139..694L}.
\cite{2011A&A...530A..95L} and \cite{2013AJ....146...30K} find the majority of regular SN~Ibc explosions to also occur in regions where the latest SF episode occurred 5-8 Myr ago, based on the H$\alpha$ EW.
A significant fraction ($\sim$20\%), however, is found in regions older than 10 Myr. 

The [\mbox{O\,{\sc iii}}] EW dependence on age is complex and depends strongly on the escape fraction of ionizing photons and on metallicity. At very low metallicity (below 1\% solar) there is not enough oxygen to produce strong lines, while stars that are very metal-enriched do not produce enough ionizing photons. \cite{2011MNRAS.415.2920I} finds that the highest EWs are attained at $\sim$0.2 $Z_{\sun}$, exactly the metallicity regime where most of our SLSNe are found. The same metallicity dependence is found by \cite{2013Natur.502..524F} who are based on \cite{2003MNRAS.344.1000B} models. 
In this metallicity regime, [\mbox{O\,{\sc iii}}] EW values greater than 300 \AA\ (400 \AA) can only be associated with starbursts younger than 5 Myr (4 Myr).
Of course, this assumes a bursty SFH but, as explained above, this is appropriate for dwarf galaxies. 
These stellar lifetimes can imply progenitors with initial masses well exceeding 60~$M_{\sun}$ \citep[e.g.][]{2002RvMP...74.1015W,MeyMae2005}.
For the host of PTF12dam, which is one of the most extreme galaxies in our sample, \cite{2014arXiv1411.1104T} are able to place constraints on the age of the most recent starburst, based on a number of age indicators. The most constraining value points towards an age of $\sim$3 Myr (120 $M_{\sun}$) supporting the idea that (at least some) SLSNe likely come from very young and very massive stars. 
The locations of the SLSN-I/R hosts on the BPT diagram (Fig.~\ref{fig:BPT}) suggest high effective temperatures in their environments \citep[probably $>50000$~K;][]{2013ApJ...766...91J,2014ApJ...795..165S}.
One way to reach these temperatures is by the contribution of massive O stars, especially if these are evolving homogeneously at low metallicity \citep{2011A&A...530A.115B,2013ApJ...766...91J}.

It has been widely suggested that H-poor SLSNe are powered by the spin-down of a rapidly rotating magnetar \citep{2010ApJ...717..245K, 2013ApJ...770..128I, 2013Natur.502..346N, 2014ApJ...787..138L}, although the modulations seen in the light curves of individual events \citep[e.g.][]{2012A&A...541A.129L,2014MNRAS.444.2096N} cannot be explained by this model.
Among the models proposed for H-poor SLSNe, the magnetar model is the only model that only requires a moderate progenitor mass \citep{2013Natur.502..346N}. 
Other models, including CSM interaction and, especially pair-instability, require a considerably more massive progenitor, as it can be deduced by derived ejected, CSM and $^{56}$Ni masses  \citep{2013ApJ...773...76C,2014arXiv1409.7728C}.
Ways to explain intensive mass loss prior to the explosion of a very massive star, and the formation of a massive CSM shell surrounding it, have recently been proposed \citep{2007Natur.450..390W,2012MNRAS.423L..92Q}.
It has also been shown that the shells forming around fast-rotating stars evolving at low metallicity can be H-poor \citep{2012ApJ...760..154C}.
Contrary to \cite{2014ApJ...787..138L}, we do not think that  arguments derived purely from environmental studies, and based on the present findings, add support to the magnetar interpretation of SLSNe-I/R. H-poor CSM interaction provides a viable alternative that is also compatible with the observed environments of H-poor SLSNe.

We propose that it should be investigated whether the extreme conditions in the birthplaces of H-poor SLSNe influence their extreme properties.
In particular it should be investigated whether the hard radiation field can directly influence the CSM interaction and thereby the SN light curves and spectra. 
Such an investigation is beyond the scope of the this paper.

After this manuscript appeared, \cite{2014arXiv1411.1060L} presented a resolved study of H-poor SLSN environments with \textit{HST}. Following a method similar to \cite{F06} they argue that H-poor SLSNe appear to be less associated than GRBs with their host galaxy UV light. They therefore disfavour our suggestion that H-poor SLSNe are among the first stars to explode in a starburst but propose instead that SLSN-I/R progenitors are less massive than those of GRBs. The trend they present (the statistical significance is low; $p=0.25$) is not at odds with our interpretation because UV light does not probe star formation at the same time-scales as H$\alpha$. UV light probes recent (16-100 Myr) star formation, while H$\alpha$ is sensitive to the ongoing SF over shorter time-scales \citep[e.g.][]{2012MNRAS.424.1372A}.
Although a correlation between the two does exist, the weak association with UV light cannot be used to rule out a very young star forming episode. 
An illustrative example to demonstrate  this,  is the case of SN~1999as. Situated at a very large projected distance  from its host galaxy (10.7~kpc; Schulze et al., in prep), it is located in a region of very low surface brightness. However, placing the slit through the exact explosion location we uncovered very strong line emission, indicative of a young starburst and of star formation that was otherwise hidden (Fig.~\ref{fig:spec1}).

\cite{2014ApJ...797...24V} also find differences between the local environments of H-poor SLSNe and those of GRBs: the \mbox{Mg\,{\sc ii}}  absorbers towards H-poor SLSNe  appear weaker than those towards GRBs. This could be explained by a combination of SLSN hosts being typically less massive, and therefore containing less gas on average, and by the gas being more ionized than in GRB hosts.

\subsubsection{Implications for the host galaxies}
\label{sec:ImplGal}

The preference of SLSNe to explode in star bursting dwarf galaxies over regular galaxies and their likely association with very massive stars, might indicate a bottom-light\footnote{we are using the term bottom-light rather than top-heavy, although the difference is mostly semantic. Top-heavy usually refers to an IMF with a slope that is less steep than the canonical value, while bottom-light usually means that low-mass stars are suppressed, i.e. there is a low-mass cut-off, independent of the slope \citep[e.g.][]{2008MNRAS.385..147D}.}
IMF in these systems or during starburst events.    
Although such an IMF is not required, it could also help explain the enhanced number of ionizing photons in H-poor SLSN environments and their position in the BPT diagram.
This is not the first time that a bottom-light IMF has been suggested for starburst environments or compact dwarf galaxies \citep{2005MNRAS.363L..31N,2007MNRAS.375..673K,2009MNRAS.394.1529D,2011MNRAS.412..979W}, but it is a completely independent way to reach this conclusion. 
The possible variation of the IMF with the environment is a topic that has not been settled \citep[see][and references therein]{2010ARA&A..48..339B}, so it is important to address this problem from many different angles. 
SN-related arguments have been used previously to infer conclusions regarding the IMF:  \cite{2010ApJ...717..342H} have suggested that the IMF in disturbed and interacting galaxies is top-heavy in order to explain the increased rate of SNe~Ibc compared to SNe~IIP in these systems. 
In fact, the conclusion reached here might be very similar as EELGs often demonstrate disturbed or interacting morphologies \citep[e.g.][]{2014arXiv1403.3441A} and so do SLSN-I/R hosts \cite[see][]{2014arXiv1411.1060L}. 
In addition, it is possibly hard to reconcile the presence of so massive stars with a regular IMF and with the limited mass in the gas reservoir of these galaxies,
especially considering the relatively low masses of the hosts.

A potentially important implication, especially for the modelling of these galaxies, is the feedback returned by these SNe.  
The kinetic energy estimates for these explosions are model-dependent. 
\cite{2013ApJ...779...98H} argue that the values reach 10$^{52}$ erg, 
although other references give more modest estimates of 2-5 10$^{51}$ erg \citep{2013ApJ...773...76C}.
In any case, it is possible that the energy output of these explosions is significantly higher than the canonical value of 10$^{51}$ erg used in galaxy simulations \citep[e.g.][]{2006MNRAS.373.1074S}.
Even if the rate of SLSN is overall low \citep{2013MNRAS.431..912Q}, their preference for a specific type of galaxies might imply that their feedback effect in these systems is worth investigating.

\subsubsection{A new way to discover and study EELGs}
\label{subsub:newway}

The high percentage of EELGs among SLSN hosts ($\sim$50\%), indicates that this can be an efficient way of detecting these rare galaxies at distances beyond the local Universe. 
Distant EELGs have hitherto been discovered  with a variety of  methods, including deep spectroscopic surveys \citep[e.g.][]{2011ApJ...743..121A,2014arXiv1403.3441A,2014A&A...568L...8A}, narrow-band imaging \citep{2013MNRAS.428.1128S} or through colour-selection criteria \citep{2009MNRAS.399.1191C,2011ApJ...742..111V}. 

Discovering EELGs through a SLSN presents a method complementary to the above but it can also offer some additional advantages both in the detection and the study of these galaxies. First, this method is completely independent of the host galaxy magnitude and it was demonstrated that it can probe also galaxies of lower luminosity and, therefore, lower mass. In fact, except from the galaxies presented in this paper, there exist SLSN hosts that are fainter (reaching down to $M_B = -13$) and that would evade detection even in moderately deep imaging surveys ($R\sim26$~mag; Schulze et al. in prep.). Spectroscopy for these objects will be expensive but will allow us to probe star formation at the lowest mass scales. 
In addition, spectroscopy of the SLSN gives the opportunity to study the host environment with UV absorption spectroscopy \citep{2012ApJ...755L..29B,2014ApJ...797...24V} as long as the redshift is high enough ($z>0.5$) so that the relevant UV absorption lines (H, Mg, Si, Fe, Zn) are shifted to the optical or NIR. 
Absorption spectroscopy probes metallicity in a way complementary to nebular emission lines but also opens  an independent window to measure the column densities, molecular and dust content and kinematics of EELGs. This method has been used successfully with GRBs \citep[e.g.][]{2007ApJ...666..267P,2009A&A...506..661L}.

The number of SLSNe remains relatively small and their redshift distribution depends on the discovery survey  \citep[e.g.][]{2011Natur.474..487Q,2014ApJ...787..138L,2013ApJ...779...98H}.
However, in the near-future, the Large Synoptic Survey Telescope (LSST) will significantly increase the number of SLSN discoveries and their redshift interval, making them potentially useful tools for cosmology \citep{2014ApJ...796...87I}.
In addition, using innovative methods, \cite{2012Natur.491..228C} have demonstrated that SLSNe can be observed up to $z = 4$, while theoretical studies extend these predictions to much higher redshifts \citep[e.g.][]{2013ApJ...777..110W}.
Therefore, large numbers of SLSN hosts (and therefore SLSN-selected EELGs) will be discovered, allowing for the formation of significant samples, redshift evolution studies and comparison with other selection methods. As demonstrated in Section~\ref{subsec:dirmet}, among these galaxies there will also be a number of very metal-poor systems. These will be ideal laboratories to study massive star-formation and chemical evolution under nearly pristine conditions \citep{2000A&ARv..10....1K}.

\subsection{Potential biases}
\label{subsec:bias}

Several biases might affect our conclusions. 
We identify and discuss the SLSN discovery selection bias, our target selection bias and the redshift bias.
Of those, the most significant that we can identify is the selection method of SLSNe that are discovered exclusively in the optical wavelengths.
An educative example for this comes from GRBs. \cite{2009ApJS..185..526F} have shown that selecting GRBs through their optical afterglow creates a bias against the more dusty sight lines. `Dark' GRBs \citep[e.g.][]{2004ApJ...617L..21J} are GRBs with a faint (or no) optical afterglow with respect to X-rays, and a big fraction of those are attributed to dust extinction.
The inclusion of dark GRBs in host galaxy studies has revealed a significant number of redder, more massive, more metal-rich and more dusty hosts   
\citep{2011A&A...534A.108K,2013ApJ...778..128P}. 
As with GRBs, it is possible that we are missing SLSNe in dusty environments, exactly because they are extincted and more difficult to detect. If `dark' SLSNe explosions exist, they might indeed revise our host galaxy population.
A claim for an obscured SLSN has been made for SN~2007va \citep{2010ApJ...722.1624K}. However, the SN nature of this event is not entirely clear and optical observations are lacking. Besides, the specific host is indeed of low luminosity and metallicity and would thus not change the paradigm presented here. 
However, \cite{2013ApJ...778..128P} have shown that there still is a dearth of massive GRB hosts at low redshifts ($z<1.5$) where our study is made, so this might also turn out to be the case for SLSNe.
In addition, the `literature' GRB sample that we are using is largely pre-\textit{Swift} and mostly selected in the optical. 
Therefore, if biased at all, it is biased towards \textit{less} massive galaxies.
This suggests that the real difference between GRB hosts and the present SUSHIES H-poor SLSN sample might be even larger.
Finally, TOUGH \citep{2012ApJ...756..187H} detected all $z<1$ GRB hosts, creating an unbiased and complete sample. 
These galaxies are still more luminous than H-poor SLSN hosts (the full SUSHIES sample; Schulze et al., in prep.).

It does not appear that this optical selection bias problem will be solved anytime soon. 
H-poor SLSNe have not been detected in the X-rays \citep[with one possible exception;][]{2013ApJ...771..136L} or the radio and, therefore, the optical seems the only realistic discovery channel. 
Furthermore, it is not clear whether the different (optical) surveys select SLSNe in different ways and how these differences affect the sample \citep[for an extensive discussion see][]{2014ApJ...787..138L}. 
Until this problem is solved, our conclusions will concern the hitherto \textit{known} population of SLSNe.
   
The `spectroscopic' sample studied in this paper did not include the hosts of all known SLSNe. This constitutes our target selection bias.
Some hosts were simply not included because of a combination of observing constraints (visibility) and available telescope time.
Most importantly, a few SLSN hosts (e.g. PTF09atu, SN~2008es) were not observed because they were found to be too faint ($R>26$ mag) and spectroscopy was very expensive. It is very likely that this \textit{does} create a bias between the spectroscopic and the overall SUSHIES sample. However, this bias would work towards extending the SLSN  host mass distribution towards lower, not higher, masses and would therefore bring it even further away from the regular CC SN and GRB host distribution.
We are presently extending our spectroscopic observations to cover both galaxies that were not observed and fainter and more distant galaxies.

The redshift bias does not appear very significant. As shown in Section~\ref{subsec:compsampl}, all samples used in this paper are at comparable redshifts (with the exception of CC SNe).
Indicatively, the sSFR for the `main sequence' of galaxies increases only by a factor of 1.85 between $z=0$ and $z=0.3$ and a factor of 1.72 between $z=0.3$ and $0.6$ \citep{2011A&A...533A.119E}.
These differences correspond to systematic offsets of $\sim$0.25 dex. 
Such an offset cannot reconcile the dramatic differences seen in Fig.~\ref{fig:MsSFR} between H-poor SLSN hosts and the CC SN hosts and SDSS galaxies at $z=0$.   
At the same time, accounting for this offset would even reduce the effective difference  between H-poor SLSNe and the VUDS EELGs (which are found at higher $z$). 
At the same redshift interval there is also no significant average metallicity evolution \citep[e.g.][]{2005ApJ...635..260S}, able to explain the observed difference.
 

\section{Summary and conclusions}
\label{sec:conc}

We have presented SUSHIES, a project to study the environments of SLSNe.
This paper focuses on SLSN hosts with spectroscopy; 
the present sample consists of 23 host galaxies, out of which 16 were observed with our own programmes and seven were retrieved from archives.
We measure emission line fluxes and equivalent widths and compute metallicities, flux ratios, ionization parameters and star formation rates.
Our results are compared with those derived for other relevant galaxy samples, such as EELGs and the hosts of GRBs and CC SNe.
For all comparison samples, all properties have been re-derived consistently based on the emission line fluxes (when available). 
The main conclusions of this study are as follows: 

\begin{enumerate}
\item We do not see any significant difference between SLSN-I and SLSN-R hosts in the present sample and for the properties examined here. For this reason, we group them together in H-poor SLSNe.
\item H-poor SLSNe explode in low-mass, metal-poor galaxies with high sSFR (median values in the spectroscopic sample: $10^{8.0}~M_{\sun}$, 0.27~$Z_{\sun}$, $10^{-8.8}M_{\sun}$~yr$^{-1}$).
\item Our H-poor sample also includes a number (3/16) of very metal-poor galaxies (below 10\% solar) as measured directly through the detection of the auroral [\mbox{O\,{\sc iii}}] $\lambda$4363 line.
\item H-poor SLSN hosts have remarkably strong emission lines. In this sample, $\sim$$50 \%$ have [\mbox{O\,{\sc iii}}] EW $>$ 100 \AA, with a median of 190 \AA\  and values reaching 800 \AA.  This cannot be a chance coincidence.
\item H-poor SLSN hosts show strong similarities with EELGs and share many common properties with these galaxies.
\item GRBs explode, on average, in less extreme and higher mass galaxies than H-poor SLSNe. 
\item We suggest that H-poor SLSNe are the result of the first stellar explosions in a starburst and that they occur earlier, on average, than GRBs. This indicates that they probably result from very massive stars. 
\item These findings could possibly support a bottom-light IMF in starburst environments. It should be investigated whether the feedback returned by these explosions affect the evolution and modelling of these systems. 
\item H-poor SLSNe present a novel method of selecting EELGs independent of their luminosity. They offer a way to probe star formation at the faint end of the galaxy luminosity function. This will become particularly interesting in the near future in view of LSST that will discover large samples of SLSNe  over a wide redshift range. 
\item We have also (for the first time) presented a systematic study of SLSN-II (H-rich) hosts. These occur in different environments than H-poor SLSNe: more massive, more metal-rich and with softer radiation fields. This suggests that the progenitors of these systems are different and that it is unlikely that SLSNe-II are just SLSNe-I surrounded by some additional H-rich CSM.  
\end{enumerate}

These conclusions will be tested in the future as SUSHIES continues to obtain data for SLSN hosts.

\section*{Acknowledgments}
We are grateful to Enrique P{\'e}rez-Montero for sharing with us his photoionization models.
We also thank: Manos Chatzopoulos for the spectrum of SN~2008am, 
Andrew Levan for discussions concerning observation planning, 
Robert Quimby for providing us blind offsets for PTF09cwl and for confirming to us the coordinates and nature of PTF10qaf, 
and Paul Vreeswijk for comments.
We acknowledge useful discussions with Jesus Zavala.
SS acknowledges support from CONICYT-Chile FONDECYT 3140534, Basal-CATA PFB-06/2007, and Project IC120009 "Millennium Institute of Astrophysics (MAS)" of Iniciativa Cient\'{\i}fica Milenio del Ministerio de Econom\'{\i}a, Fomento y Turismo.
The research activity of AdUP, CT and JG is supported by Spanish research project AYA2012-5 39362-C02-02. AdUP acknowledges support by the European Commission under the Marie Curie Career Integration Grant programme (FP7-PEOPLE-2012-CIG 322307).
FEB acknowledges support from Basal-CATA PFB-06/2007, CONICYT-Chile (FONDECYT 1141218, "EMBIGGEN" Anillo ACT1101) and 
Project IC120009 "Millennium Institute of Astrophysics (MAS)" of the Iniciativa Cient\'{\i}fica Milenio del Ministerio de Econom\'{\i}a, Fomento y Turismo.
AG acknowledges financial support from the European Union Seventh Framework Programme (FP7/2007-2013) under grant agreement n. 267251 (AstroFIt).
KGH acknowledges support provided by the National Astronomical Observatory of Japan as Subaru Astronomical Research Fellow, and the Polish National Science center grant 2011/03/N/ST9/01819.
EI acknowledges funding from CONICYT/FONDECYT postdoctoral project N$^\circ$:3130504.
JV is supported by Hungarian OTKA Grant NN 107637.
This work was supported in part by NSF grant AST 11-09801 to JCW and by NSF grant  PHYS-1066293 and the hospitality of the Aspen Center for Physics extended to JCW.
The Dark Cosmology Centre is funded by the Danish National Research Foundation. 
This research has made use of the GHostS data base, which is partly funded by \textit{Spitzer}/NASA grant RSA Agreement No. 1287913.
Based on observations made with the Gran Telescopio Canarias (GTC), installed in the Spanish Observatorio del Roque de los Muchachos of the Instituto de Astrof\'{i}sica de Canarias, in the island of La Palma.
Based on observations made with Magellan as part of CN2013A-195, CN2013B-70, CN2013B-34, CN2014A-114.
Based on observations made with ESO Telescopes at the La Silla Paranal Observatory under programme IDs 089.D-0902(ABC), 091.D-0734(ABC), 290.D-5139(A), 093.D-0867(A).

\bibliographystyle{aa}  

\bibliography{SLSNhosts1.bib}


\appendix

\section{Notes on individual objects}

\subsection{PTF10vqv}

For this galaxy we used direct integration to measure the line fluxes, as Gaussian profiles did not provide accurate fits at the resolution of X-shooter. This could possibly indicate multiple kinematic components in this host.

\subsection{PS1-10bzj}

We were not able to satisfactorily match the multi-band photometry for this galaxy provided by \cite{2013ApJ...771...97L} by scaling and warping the host galaxy spectrum and by doing synthetic photometry. For this reason, we finally did not apply any photometry scaling correction. This might explain some discrepancies with values reported in this paper. 

\subsection{PTF11dsf}
\label{app:11dsf}

The observations of this galaxy contain a few peculiarities. Our photometry (Table~\ref{tab:res} and Schulze et al. in prep.) is quite discrepant from the one obtained by SDSS (it is brighter than in SDSS). In addition, the spectrum shows a broad component under both the H$\alpha$ and the [\mbox{O\,{\sc iii}}] line (Fig.~\ref{fig:spec3}).  
This can possibly indicate a contribution from a persisting SN or from AGN activity.
Such a broad component is also rarely seen in star forming galaxies and can be attributed to strong outflows caused by SN winds \citep{2012ApJ...754L..22A}.
Since this is a SLSN-II, the contribution of a long-lasting SN two years after discovery can not be excluded.  
We note that if SN flux contaminates our measurements, this will only increase the observed difference of H-rich to H-poor SLSN hosts, as the host of PTF11dsf was the one with the largest similarities to H-poor SLSN hosts.

\subsection{PTF12dam}

We have adopted a host galaxy extinction from the ratio of H$\gamma$ to H$\beta$ (and not from H$\alpha$ to H$\beta$ as done with all other galaxies for which H$\alpha$ is available). This was done because both H$\gamma$ and H$\beta$ have sufficient S/N and they are found on the side of the spectrum obtained with the same (bluer) grism. This explains a small discrepancy with the value adopted by \cite{2014arXiv1411.1104T} who are (partially) based on the same data.

%
%

\section{Individual spectra and line fluxes}

\begin{table*}
 \centering
 \begin{minipage}{172mm}
  \caption{Line fluxes, corrected for foreground Galactic extinction. Units: $10^{-17}$ erg~s$^{-1}$~cm$^{-2}$. Upper limits are 2$\sigma$. The SNe are listed by discovery year. For three SNe we provide measurements at two different locations.}
    \label{tab:fluxes}
  \begin{tabular}{@{}lrcccccc@{}}
\hline
SLSN host & [\mbox{O\,{\sc ii}}] $\lambda\lambda$3727,29 & [\mbox{O\,{\sc iii}}] $\lambda$4363  & H$\beta$ & [\mbox{O\,{\sc iii}}] $\lambda$4959  & [\mbox{O\,{\sc iii}}] $\lambda$5007  & H$\alpha$ & [\mbox{N\,{\sc ii}}] $\lambda$6584   \\ 
\hline
SN 1999as &  29.81 (2.81) & \ldots &   3.26 (1.46) &   3.65 (0.65) &  10.95 (1.94) &  17.96 (1.41) &   4.88 (0.58) \\ 
-- SN location &   $<$   7.96  & \ldots &   3.14 (0.49) &   4.59 (0.33) &  13.76 (0.99) &  11.57 (0.81) &   $<$   1.21  \\ 
SN 1999bd &  82.53 (3.42) & \ldots &  24.26 (3.12) &  12.63 (3.17) &  35.57 (3.49) & 112.75 (3.20) &  34.92 (1.68) \\ 
SN 2006oz &   3.34 (0.39) & \ldots &   0.91 (0.36) &   0.98 (0.35) &   3.52 (0.40) &   3.13 (0.32) &   $<$   1.17  \\ 
SN 2006tf &  45.15 (2.91) & \ldots &  11.12 (1.24) &  11.23 (1.42) &  30.02 (1.40) &  49.02 (1.37) &   4.62 (0.66) \\ 
SNLS 06D4eu &   3.74 (0.17) & \ldots &   1.60 (0.22) &   1.98 (0.23) &   6.09 (0.30) &   3.72 (0.18) &   0.38 (0.07) \\ 
SN 2007bi &   5.73 (0.43) & \ldots &   1.78 (0.26) &   1.86 (0.09) &   5.57 (0.27) &   5.82 (0.33) &   0.38 (0.13) \\ 
SN 2008am &  57.64 (3.17) & \ldots &  18.47 (1.95) &  14.31 (1.85) &  41.05 (2.78) &  75.01 (3.95) &  10.31 (0.95) \\ 
SN 2009jh & \ldots & \ldots & \ldots & \ldots &   $<$   0.30  &   $<$   0.30  & \ldots \\ 
PTF09cnd &  16.17 (0.75) & \ldots &   3.92 (0.42) &   4.40 (0.48) &  13.39 (0.68) &  13.50 (0.14) &   1.11 (0.34) \\ 
SN 2010gx &  18.37 (1.02) &   3.29 (0.40) &  15.39 (0.58) &  25.35 (0.54) &  75.64 (0.64) &  57.07 (0.40) &   0.88 (0.14) \\ 
SN 2010kd &  15.97 (2.66) &   2.52 (0.78) &  15.86 (0.85) &  25.48 (0.82) &  76.99 (0.95) &  48.52 (0.68) &   1.44 (0.37) \\ 
PTF10hgi &   4.47 (0.44) & \ldots &   0.56 (0.23) &   0.87 (0.21) &   2.86 (0.23) &   3.97 (0.18) &   0.61 (0.11) \\ 
PS1-10bzj &  11.96 (0.68) &   0.83 (0.24) &  12.30 (0.76) &  25.31 (0.60) &  74.38 (0.82) & \ldots & \ldots \\ 
PTF10heh &   6.60 (0.24) & \ldots &   3.29 (0.33) &   2.89 (0.37) &   8.89 (0.44) &   7.96 (0.30) &   1.01 (0.11) \\ 
PTF10qaf &  67.74 (3.46) & \ldots &  37.58 (4.45) &   7.61 (1.71) &  22.82 (5.12) & 138.47 (3.16) &  55.34 (1.50) \\ 
 -- SN location &  19.50 (1.59) & \ldots &   7.15 (2.99) &   4.21 (1.25) &  12.63 (3.74) &  14.43 (1.58) &   3.09 (1.02) \\ 
PTF10vqv &   8.27 (0.06) &   0.59 (0.18) &   6.29 (0.03) &   9.87 (0.02) &  30.04 (0.03) &  11.84 (0.03) &   $<$   0.48  \\ 
SN 2011ke &  49.49 (1.46) &   5.72 (0.39) &  50.61 (0.85) &  70.08 (0.94) & 213.88 (1.49) & 148.02 (1.14) &   $<$   1.65  \\ 
 -- tadpole tail &   9.37 (0.42) & \ldots &   1.85 (0.12) &   2.41 (0.14) &   6.24 (0.28) &   6.65 (0.21) &   $<$   0.68  \\ 
SN 2011kf &   4.99 (1.11) & \ldots &   2.69 (0.32) &   4.24 (0.54) &  12.44 (0.63) &   9.61 (0.46) &   $<$   1.02  \\ 
PS1-11ap &  11.15 (0.85) & \ldots &   3.69 (0.45) &   4.42 (0.56) &  13.26 (1.69) & \ldots & \ldots \\ 
PTF11dsf &  93.96 (3.98) & \ldots &  35.08 (1.96) &  35.89 (2.70) & 135.49 (2.61) & 102.35 (3.01) &   $<$   5.94  \\ 
SN 2012il &  35.71 (1.80) &   3.89 (0.76) &  28.00 (1.80) &  54.29 (1.96) & 161.05 (2.64) &  83.77 (2.34) &   $<$   3.05  \\ 
PTF12dam & 1299.56 (4.78) &  70.59 (0.85) & 842.95 (1.76) & 1667.82 (2.40) & 5006.93 (4.05) & 3006.39 (2.19) & 102.43 (0.35) \\ 
SSS120810 &  14.80 (3.07) & \ldots &   5.51 (1.29) &   3.57 (0.74) &  12.28 (1.38) &  16.67 (1.03) &   $<$   1.31  \\ 
\hline 
\end{tabular}
\end{minipage}
\end{table*}

\clearpage

\begin{figure*}
\includegraphics[width=\textwidth]{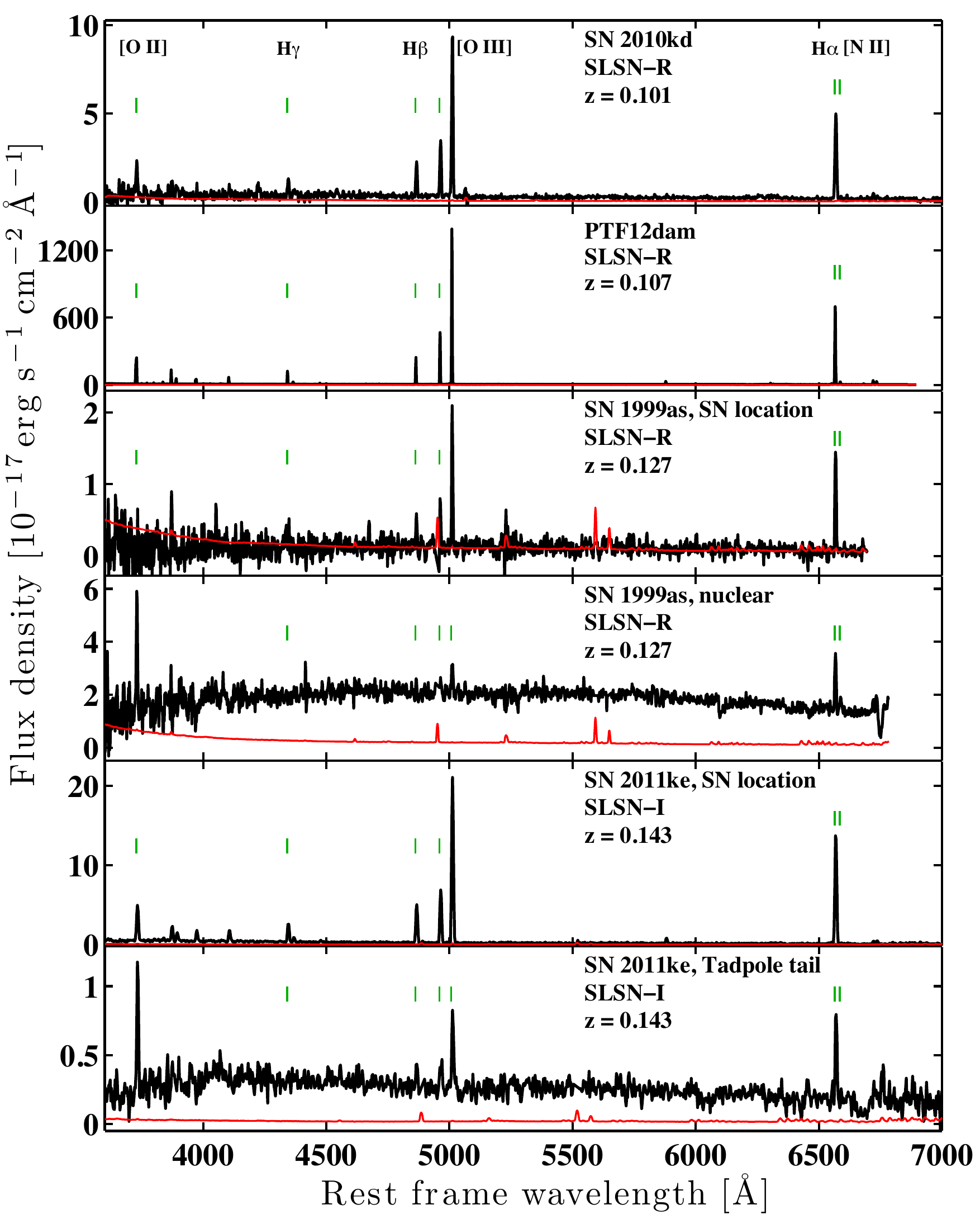}
\caption{Low-resolution spectra of SLSN-I and SLSN-R hosts with increasing redshift. 
The main emission lines are labeled in the top panel.
Green ticks mark their locations in every panel (except when they are very strong).
The error spectrum is shown in red.
Many skyline residuals have been removed manually for presentation purposes. 
For SN~1999as and SN~2011ke  we show spectra both at the exact explosion location and at the galaxy nucleus.}
\label{fig:spec1}
\end{figure*}

\clearpage

\begin{figure*}
\includegraphics[width=\textwidth]{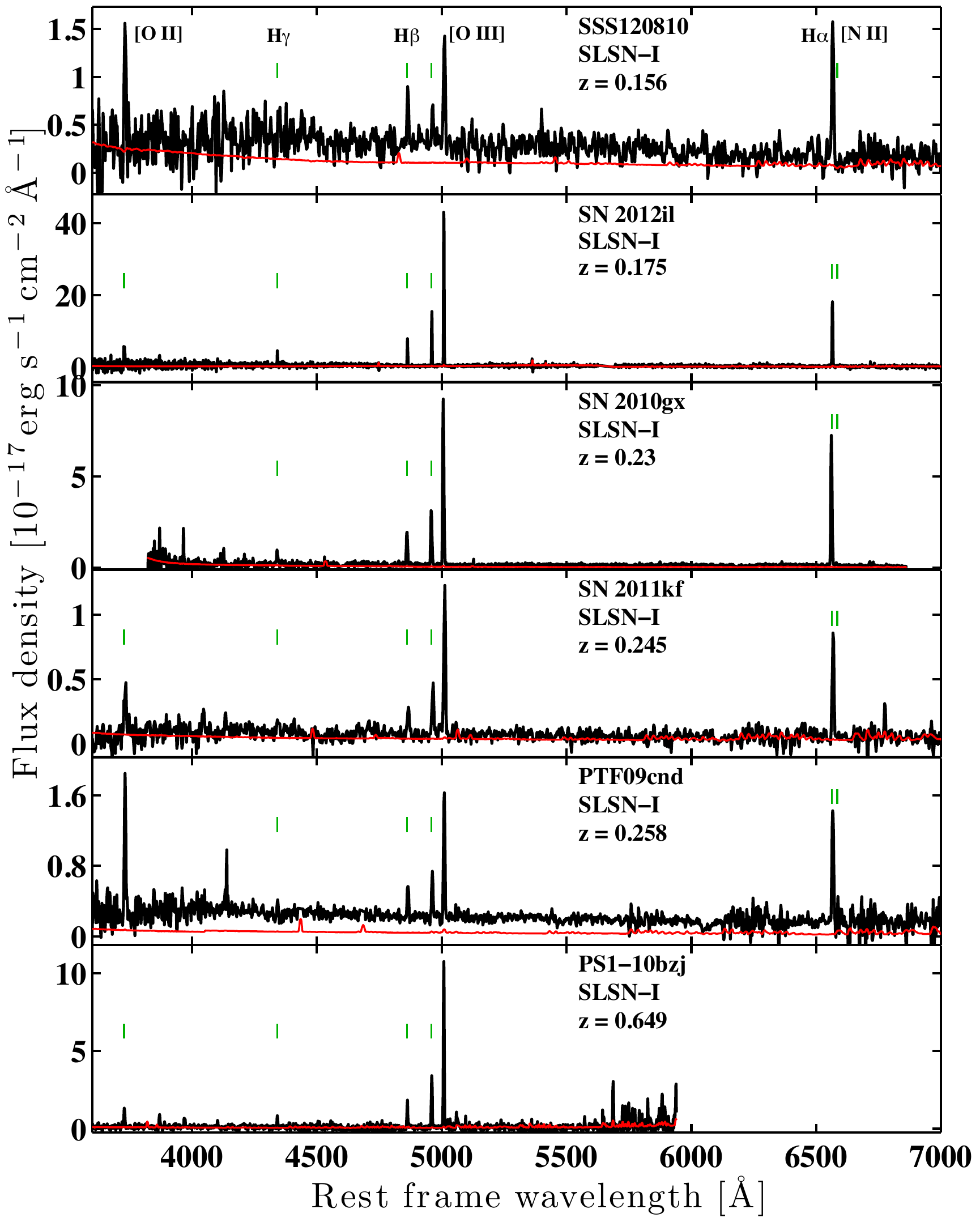}
\caption{More low-resolution spectra of H-poor SLSN hosts 
(see Fig.~\ref{fig:spec1} for a description of the lines and symbols).
In most H-poor SLSN host spectra we observe that the ratio of [\mbox{N\,{\sc ii}}] (if detected at all) to H$\alpha$ is low, indicating low metallicity.
At the same time we observe a high [\mbox{O\,{\sc iii}}] to [\mbox{O\,{\sc ii}}] ratio, indicating high ionization.
Most H-poor SLSNe explode in galaxies that can be classified as EELGs.}
\label{fig:spec2}
\end{figure*}

\clearpage

\begin{figure*}
\includegraphics[width=\textwidth]{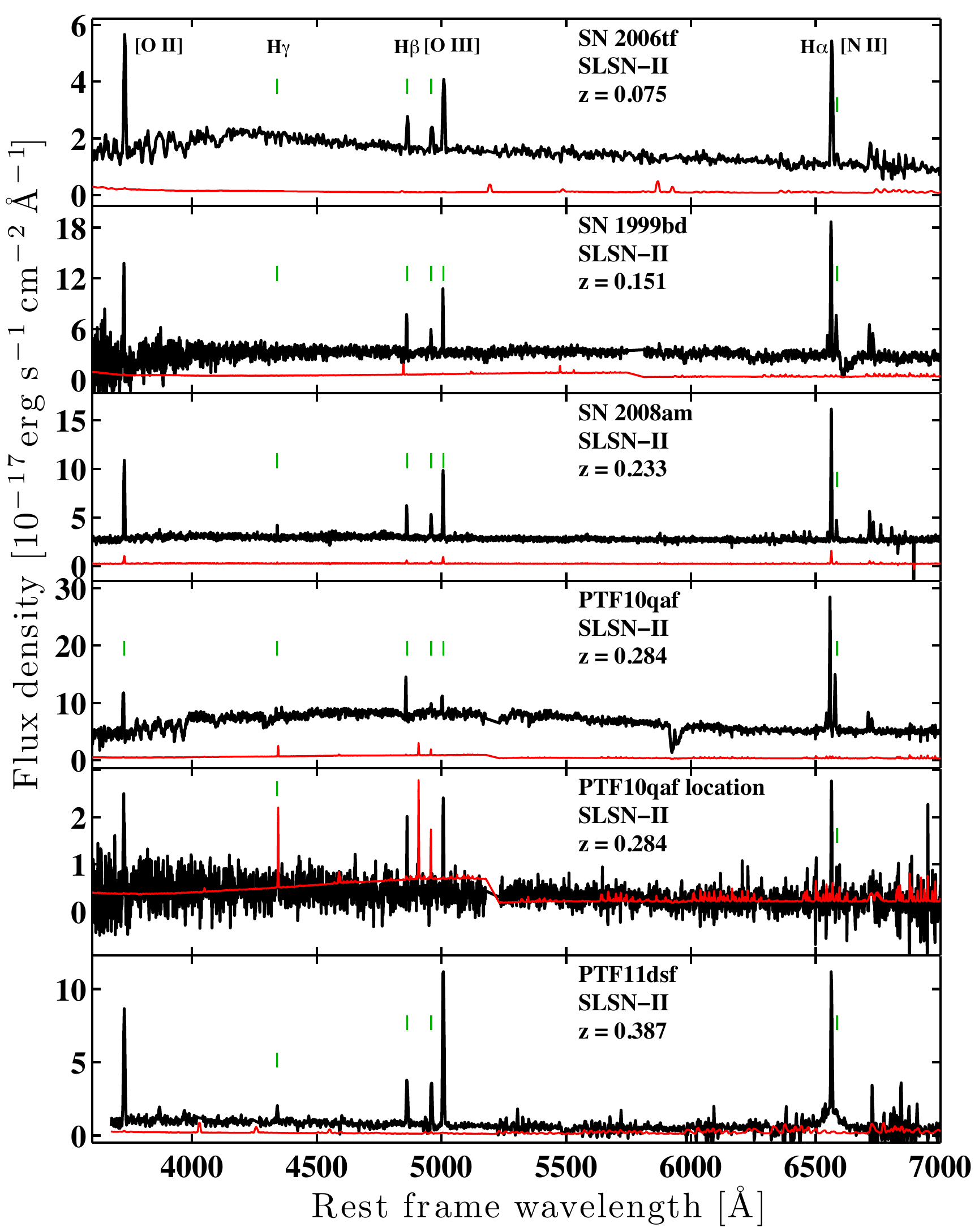} 
\caption{Low-resolution spectra of SLSN-II hosts (see Fig.~\ref{fig:spec1} for a description of the lines and symbols).
For PTF10qaf we show both a nuclear galaxy spectrum and a spectrum extracted at the SN location.
Overall, SLSN-II host spectra show a higher  [\mbox{N\,{\sc ii}}] to H$\alpha$ and a lower [\mbox{O\,{\sc iii}}] to [\mbox{O\,{\sc ii}}] ratio than H-poor SLSN hosts.
PTF11dsf demonstrates a broad base on the H$\alpha$ line (see discussion in Appendix~\ref{app:11dsf}).
}
\label{fig:spec3}
\end{figure*}

\begin{figure*}
\includegraphics[width=\textwidth]{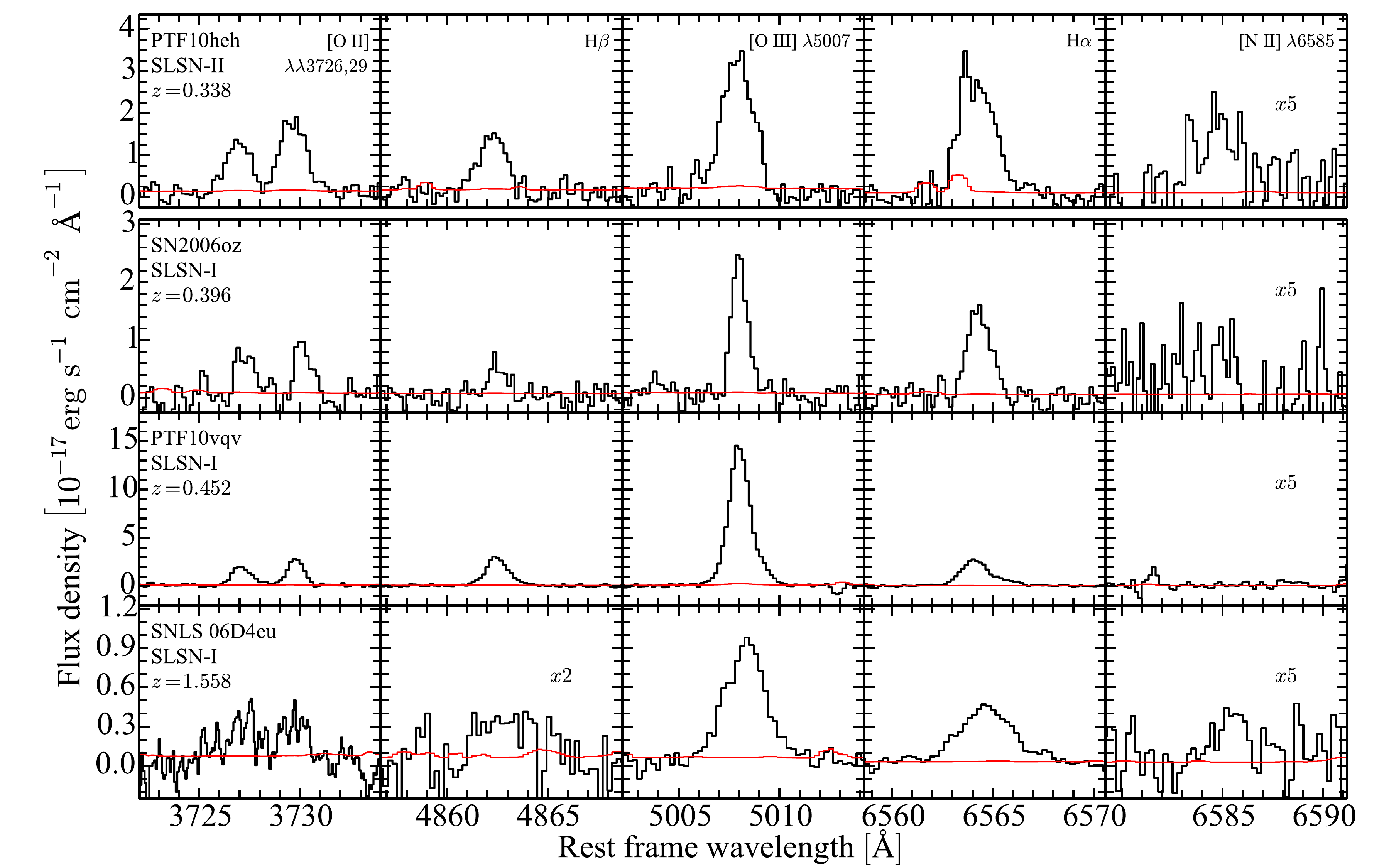} 
\caption{X-shooter medium resolution spectra of SLSN hosts.
We only show the portions containing the most important emission lines (labeled in the top panels).
The error spectrum is shown in red.
For presentation purposes the flux density of a few lines (all [\mbox{N\,{\sc ii}}] and a single  H$\beta$  line) has been multiplied by a constant factor as indicated in the corresponding panels.
SN~2009jh is not shown as we do not detect neither continuum nor line emission.
}
\label{fig:spec4}
\end{figure*}

\label{lastpage}

\end{document}